\documentclass[10pt,journal,compsoc]{IEEEtran}

\ifCLASSOPTIONcompsoc
  \usepackage[nocompress]{cite}
\else
  \usepackage{cite}
\fi

\ifCLASSINFOpdf
\else
\fi
\DeclareUnicodeCharacter{0301}{\'{e}}
\UseRawInputEncoding
\usepackage{cite}
\usepackage{amsmath,amssymb,amsfonts}
\usepackage{algorithmic}
\usepackage{graphicx}
\usepackage{textcomp}
\usepackage{multirow}
\usepackage{makecell}
\usepackage{xcolor}
\usepackage{graphicx}
\usepackage{subfigure}
\usepackage{graphbox}
\usepackage{framed}
\usepackage{tcolorbox}
\usepackage[justification=centering]{caption}
\usepackage{ragged2e}
\usepackage{array}
\usepackage{pifont}
\usepackage[numbers,sort&compress]{natbib}

\begin{document}
\title{Do Pre-trained Language Models Indeed Understand Software Engineering Tasks?}

\author{Yao Li,
        Tao Zhang,
        Xiapu Luo,
        Haipeng Cai,
        Sen Fang,
        Dawei Yuan
\IEEEcompsocitemizethanks{\IEEEcompsocthanksitem Y. Li, T. Zhang, S. Fang, and D. Yuan are with the School of Computer Science and Engineering, Macau University of Science and Technology, Macao, China.
\protect\\
E-mail: 2109853gia30001@student.must.edu.mo, tazhang@must.edu.mo, fangsen1996@gmail.com, wu.xiguanghua2014@gmail.com
\IEEEcompsocthanksitem  X. Luo is with the Department of Computing, Hong Kong Polytechnic University, Hong Kong, China.
\protect\\
E-mail: csxluo@comp.polyu.edu.hk

\IEEEcompsocthanksitem  H. Cai is with the School of Electrical Engineering and Computer Science, Washington State University Pullman, Washington, USA
\protect\\
E-mail: haipeng.cai@wsu.edu

\IEEEcompsocthanksitem Tao Zhang is a corresponding author.

}
\thanks{}}

\markboth{}%
{Shell \MakeLowercase{\textit{et al.}}: Bare Demo of IEEEtran.cls for Computer Society Journals}

\IEEEtitleabstractindextext{%
\justifying
\begin{abstract}
Artificial intelligence (AI) for software engineering (SE) tasks has recently achieved promising performance. In this paper, we investigate to what extent the pre-trained language model truly understands those SE tasks such as code search, code summarization, etc. We conduct a comprehensive empirical study on a board set of AI for SE (AI4SE) tasks by feeding them with variant inputs: 1) with various masking rates and 2) with sufficient input subset method.
Then, the trained models are evaluated on different SE tasks, including code search, code summarization, and duplicate bug report detection. Our experimental results show that pre-trained language models are insensitive to the given input, thus they achieve similar performance in these three SE tasks. We refer to this phenomenon as {\em overinterpretation}, where a model confidently makes a decision without salient features, or where a model finds some irrelevant relationships between the final decision and the dataset.
Our study investigates two approaches to mitigate the overinterpretation phenomenon: whole word mask strategy and ensembling.
To the best of our knowledge, we are the \textit{first} to reveal this overinterpretation phenomenon to the AI4SE community, which is an important reminder for researchers to design the input for the models and calls for necessary future work in understanding and implementing AI4SE tasks.
\end{abstract}

}

\maketitle

\IEEEdisplaynontitleabstractindextext

\IEEEpeerreviewmaketitle

\IEEEraisesectionheading{\section{Introduction}\label{sec:introduction}}

\IEEEPARstart{G}{iven} the great potential of artificial intelligence, applying AI for software engineering gains encouraging results in software quality \cite{GeziciBahar2022Systematic, HuangAiMing2014Study}, software development \cite{ChenShih-Yeh2022Exploring, Palomo-DuarteManuel2021Assessment}, and software project management \cite{SheorajYugeshwaree2022UsingAI, ZhengZhi-Ming2008Earned}. Despite early successes, AI4SE suffers fundamental explainability problems for its performance \cite{EspinozaGabrielZ2021EDLm}. The major reason is that the inside of the neural model is still as mysterious as a black box for researchers. 
To reveal the nature of AI, recent studies explore controlled empirical studies by specifically targeting on one task. 
For example, Qu et al. \cite{QuYu2021Evaluating} conduct an extensive empirical study to evaluate network embedding algorithms in bug prediction. Different from all previous empirical studies in AI4SE, our empirical study focuses on the impact of input variations on pre-trained language models (PLMs) applied to AI4SE tasks. Specifically, we use a masking strategy or a sufficient subset of inputs (SIS) algorithm to control the inputs of the model for observing the model performance. Therefore, the design of our empirical study is under the hyperthesis that different keywords (unmasked) lead the trained model to achieve different levels of performance. Surprisingly, our experimental results show that by varying masking rate from 15\% and 80\%, the neural models archive similar results. For example, with 80\% masking rate, Bidirectional Encoder Representations from Transformers (BERT) \cite{DevlinJacob2018BERT} still learns good pre-trained representations and keep more than 90\% of the performance on downstream tasks. We call this phenomenon ``overinterpretation".

\begin{table*}[]
\centering
\caption{The overview of RQs and findings.}
\begin{tabular}{m{5.5cm}m{5cm}m{6cm}}
\hline
RQ & Task \& Methodology        & Finding 
\\ \hline
Do software engineering tasks (code search, code summarization, and duplicate bug report detection) suffer from overinterpretation?  & Code search, code summarization, duplicate bug report detection: Masking \& SIS            &    Software engineering tasks suffer from overinterpretation.     
\\ \hline
Does overinterpretation depend on software engineering tasks and how prevalent is overinterpretation in PLMs?  & GPT, BERT, XLNet: Masking \& SIS           &  Overinterpretation is not dependent on software engineering tasks and is also prevalent in pre-trained language models.     
\\ \hline
What is the impact of overinterpretation? What are the challenges in mitigating overinterpretation in general and how to mitigate overinterpretation?   & Whole word masking \& Ensembling &    The main challenge is that overinterpretation is more difficult to detect. Whole word masking and ensembling can mitigate overinterpretation.     \\ \hline
\label{rq}
\end{tabular}
\end{table*}

\begin{figure}[h]
    \centering
    \includegraphics[scale=0.35]{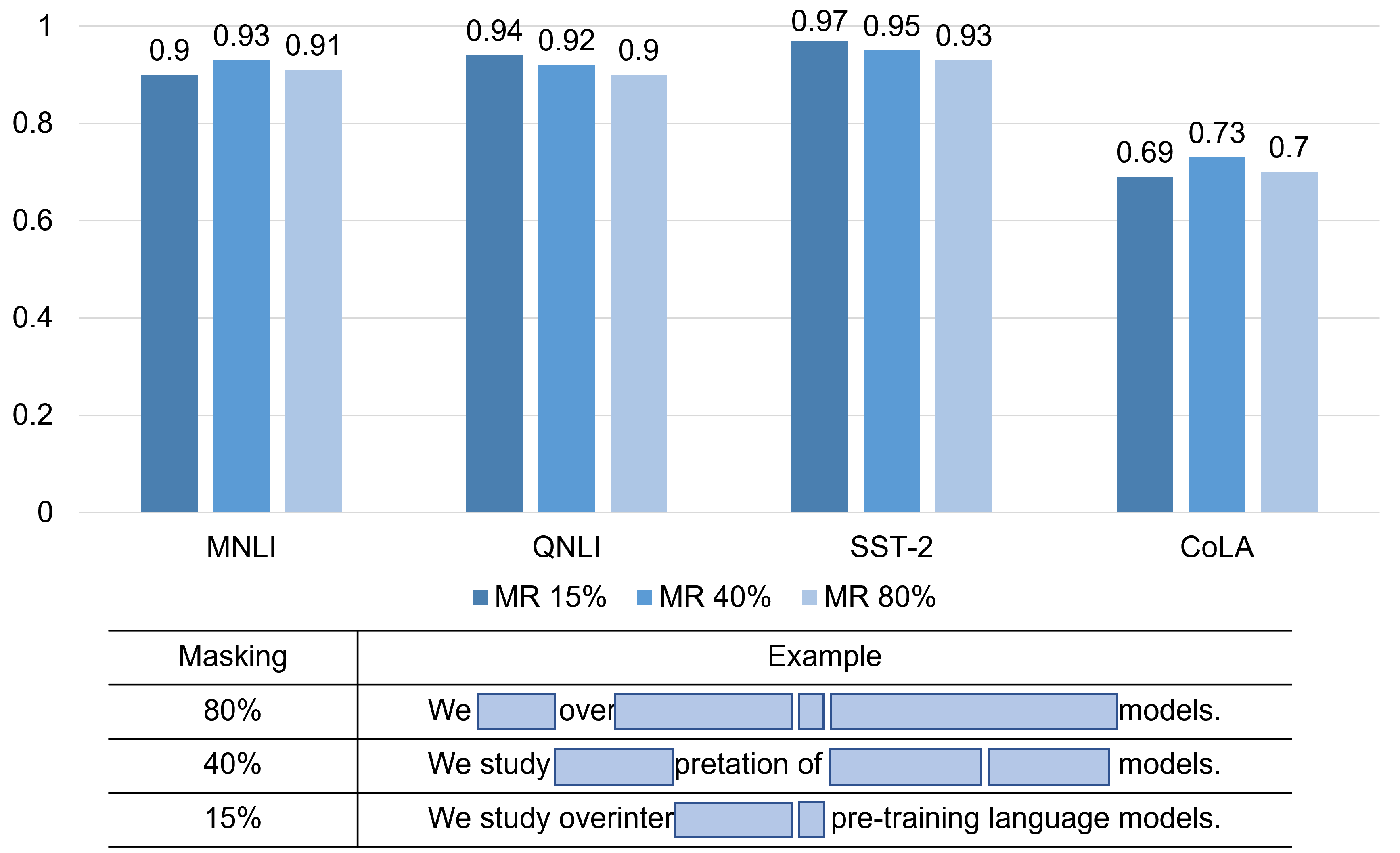}
    \caption{Performance of PLMs under different masking rates. ``MR" means masking rate. Different mask rate leads to similar performance in the four considered metrics.}
    \label{fig:text}
\end{figure}

\begin{figure}[h]
    \centering
    \includegraphics[scale=0.6]{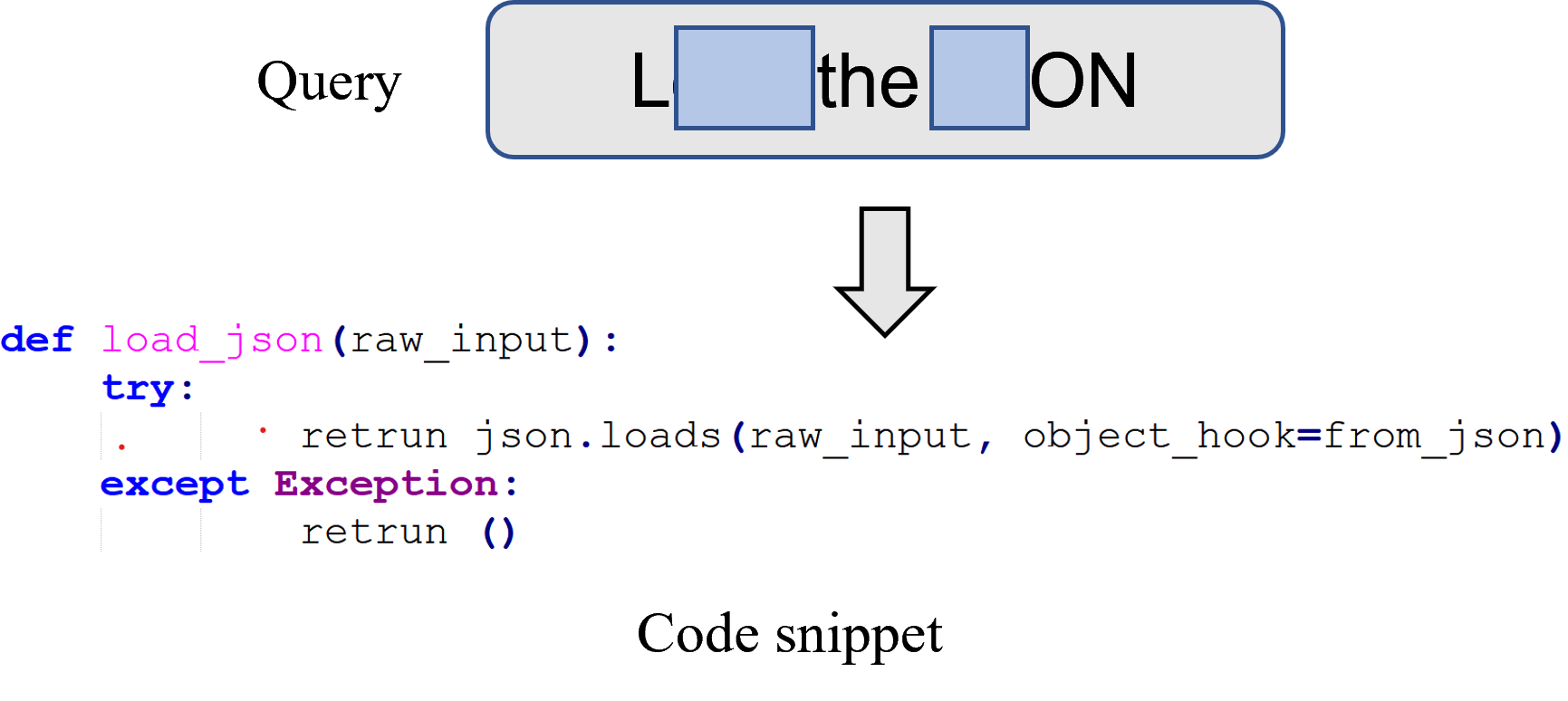}
    \caption{A code search example after the query is masked. Blue block indicates masking.}
    \label{fig:fig}
\end{figure}

Overinterpretation is a type of deep learning (DL) model failure, where a model confidently makes a decision without salient features (e.g., keywords), or where a model makes a prediction by utilizing some irrelevant relationships between the final decision and the dataset (i.e., classifying images by background pixels). However, overinterpretation can easily be misleading. Because it looks like the model can even work under bad conditions. For example, the model can classify images in which only 10\% of the pixels are retained, and the highly sparse, unmodified subsets of pixels in images suffice for image classifiers to make the same predictions as on the full images \cite{CarterBrandon2020Overinterpretation}. Meanwhile, we have found this phenomenon in SE tasks. Fig. \ref{fig:text} depicts examples of masking 15\%, 40\%, and 80\%, as well their downstream task performance. With 80\% masking rate, BERT still learns good representations and keeps more than 90\% of the performance on downstream tasks. Even in just 40\% masking rate, we can no longer understand the meaning of the sentence, but the model still makes accurate judgments. Fig. \ref{fig:fig} presents an example of SE task, i.e., code search. When the query is partially obscured, the model can still find the corresponding code snippet accurately. But from a human perspective, the masked query, with only a few letters left. It cannot be interpreted by humans, much less finding the code snippet it describes. However, these masked meaningless characters also achieve similar results as the full query. Given that the DL models in the above examples which have such remarkable success in SE tasks, it is natural to ask why they perform so well, after the inputs have been masked, what kinds of features these models are learning, and whether they can understand features, i.e., \textit{can they indeed understand SE tasks?}

To answer these questions, the key point is the input features. The features used by the model are derived from the dataset. However, dataset has implicit biases and unique statistical signals that are often introduced during the dataset generation/solidification/labeling process \cite{Chakraborty2022TSE}. These biases and statistical signals often allow DL models to achieve high accuracy in test data by learning highly specific features unique to that dataset rather than generalizable features or key features under human understanding. For example, Ribeiro et al. \cite{RibeiroMarcoTulio2016"SIT} describe an example of a classifier that classifies wolves and huskies. They find that the classier predicts ``Wolf" if there is snow (or light background at the bottom), and ``Husky" otherwise, regardless of animal color, position, pose, etc. Biases in the dataset and unique statistical signals are learned by the model and cause overinterpretation \cite{Chakraborty2022TSE}.

In this paper, we conduct the first comprehensive empirical study on the overinterpretation of PLMs applied in the SE tasks. Our study contains two parts, task-oriented and model-oriented. In the task-oriented part, we study three SE tasks, code search \cite{GuXiaodong2018DeepCS,FengZhangyin2020CodeBERT}, code summarization \cite{AhmadWasiUddin2020Transformer,IyerSrinivasan2016Summarizing}, and duplicate bug report detection \cite{XiaoGuanping2020HINDBR,BudhirajaAmar2018DWEN}. These tasks are not only widely used in software development and maintenance, but also encompass language processing techniques such as natural language (NL) to programming language (PL) translation, PL to NL translation, and NL classification. In the model-oriented part, we study three famous PLMs, Generative Pre-Training (GPT) \cite{Radford2018Corpus}, BERT \cite{DevlinJacob2018BERT}, and XLNet \cite{YangZhilin2019XLNET}. These PLMs are widely used in various tasks. The purpose of this study is to provide a systemic and generalized understanding of the overinterpretation of PLMs, which could improve model architecture and solve potential issues such as misclassification, low generalization, etc. We set three research questions (RQ) to verify the overinterpretation. The RQs and findings in the paper are shown in Table \ref{rq}. We describe them in details as following subsections.

\subsection{SE tasks analysis}

To better investigate the models in the SE tasks, we analyze three SE tasks, code search, code summarization, and duplicate bug report detection. With the support of representative studies, our analysis is driven by the following research question:

\begin{tcolorbox}[title = {RQ 1:},
  colframe = gray!30!white, colback = gray!10!white,
  colbacktitle = gray!30!white,
  coltext = black!50!black,
  coltitle = black!90!white]
  Do software engineering tasks (code search, code summarization, and duplicate bug report detection) suffer from overinterpretation?
\end{tcolorbox}

\subsubsection{Code search}

The basic principle of code search is to accept full queries and find the correct code snippets. However, our analysis results show that PLMs \cite{GuXiaodong2018DeepCS,FengZhangyin2020CodeBERT} can still find the corresponding code snippets after entering a masked query (even masking 80\% of the query) or SIS (a sparse set of unrelated characters). The masked query and SIS retain only meaningless letters or non-essential words. However, models make accurate decisions based on these confusing letters. For details, please refer to Section \ref{sec_cs}.

\subsubsection{Code summarization}

Despite multiple publications proposing new code summarization methods \cite{SridharaGiriprasad2010Tags,PollockLori2013NLSA,FowkesJaroslav2017AfSC}, we do not find an analysis of overinterpretation. Therefore, we analyze several approaches \cite{AhmadWasiUddin2020Transformer,IyerSrinivasan2016Summarizing} to investigate whether they overinterpret the dataset.

The investigation results show that, despite the lack of input code (masking strategy and SIS), these studies \cite{AhmadWasiUddin2020Transformer,IyerSrinivasan2016Summarizing} are able to accurately generate the corresponding summaries. However, the input that remains is unrelated and confusing, and cannot be understood from a human perspective. For more details, please refer to Section \ref{sec_summ}.

\subsubsection{Duplicate bug report detection}

Capturing and tagging duplicate bug reports is crucial to avoid the assignment of the same bug to different developers. We design a variety of experiments to study and analyze some studies \cite{XiaoGuanping2020HINDBR,BudhirajaAmar2018DWEN}.

We use three different masking strategies (15\% masking rate, 40\% masking rate, 80\% masking rate) and SIS to train these two models \cite{XiaoGuanping2020HINDBR,BudhirajaAmar2018DWEN} separately. When lacks most of the content of the description tag, the model can still detect the duplicate report. The retained descriptions are unreadable and meaningless, much less containing salient features. For the details, please refer to Section \ref{sec_dbr}.

\subsection{PLMs analysis}

To demonstrate that overinterpretation is not task-dependent and is prevalent in pre-trained language models, we choose three representative PLMs (GPT, BERT, and XLNet) for evaluation.

\begin{tcolorbox}[title = {RQ 2:},
  colframe = gray!30!white, colback = gray!10!white,
  colbacktitle = gray!30!white,
  coltext = black!50!black,
  coltitle = black!90!white]
  Does overinterpretation depend on software engineering tasks and how prevalent is overinterpretation in PLMs? 
\end{tcolorbox}

We find that PLMs \cite{Radford2018Corpus,DevlinJacob2018BERT,YangZhilin2019XLNET} can still make accurate decisions under conditions where most of the data and salient features are missing (by using masking strategies and SIS). Overinterpretation not only depends on SE-related tasks but is also prevalent in PLMs. For the details, please refer to Section \ref{sec_model}. 

\subsection{Impact Analysis \& Mitigation}

Overinterpretation is a potential pitfall. It suggests that the pre-trained language model learns not the salient features in the dataset, such as some keywords in the text. Instead, it learns statistical signals that are unique to the data. Thus, in this paper, we are interested in exploring what the hindrances to alleviate this flaw are and how to mitigate overinterpretation.

\begin{tcolorbox}[title = {RQ 3:},
  colframe = gray!30!white, colback = gray!10!white,
  colbacktitle = gray!30!white,
  coltext = black!50!black,
  coltitle = black!90!white]
  What is the impact of overinterpretation? What are the challenges in mitigating overinterpretation in general and how to mitigate overinterpretation?  
\end{tcolorbox}

There are three main challenges. First, overinterpretation is not well understood and studied at present. Meanwhile, overinterpretation can be misleading. Second, overinterpretation is not easily detected. Overinterpretation may come from real statistical signals in the distribution of the underlying dataset. Finally, the pre-trained language model is a black-box model. Moreover, we find two ways to mitigate overinterpretation through experiments, whole word mask and ensembling. These two methods can enrich the dataset used by the model. For the details, please refer to Section \ref{sec_miti}.

\subsection{Contributions}
In summary, this paper makes the following contributions:

\begin{itemize}
\item We perform the \textit{first} comprehensive study on the overinterpretation of pre-trained language models in SE. We demonstrate that PLMs in SE suffers from overinterpretation.
\item We design two schemes to reveal overinterpretation. One is a different masking rate scheme and the other is a sufficient input subset scheme.
\item We find two ways to mitigate overinterpretation, whole word mask strategy and ensembling. These two methods can enrich model learning to mitigate overinterpretation.
\end{itemize}

The rest of this paper is organized as follows. In Section \ref{sec_back}, we give an overview of PLMs, AI4SE tasks, and overinterpretation. Section \ref{sec_method} describes the study methodology and Section \ref{sec_setup} describes datasets and experimental setup. Section \ref{sec_task} and Section \ref{sec_model} present the analysis of the overinterpretation in SE tasks and PLMs. Section \ref{sec_miti} introduces the impact of overinterpretation and mitigation measures.
Section \ref{sec_threats} describes the threats to validity. We survey related work in Section \ref{sec_rw} and conclude this paper in Section \ref{sec_conclu}. 

{\setlength{\parindent}{0cm}
\textbf{Data Avalibility:} The data underlying this paper are available in https://doi.org/10.17632/yz3gnwvzfm.1.}

\section{Background}
\label{sec_back}

\subsection{PLMs}
Pre-training has always been an effective strategy to learn the parameters of deep neural networks, which are then fine-tuned on downstream tasks. As early as 2006, the breakthrough of deep learning came with greedy layer-wise unsupervised pre-training followed by supervised fine-tuning \cite{HintonGeoffreyE2006Fast}. In natural language processing (NLP), PLMs on large corpus have been proven to be beneficial for the downstream NLP tasks. With the development of computers, PLMs change a lot, from shallow word embedding to deep neural networks. 

Language modeling (LM) objectives for pre-training mainly fall into two categories: (1) autoregressive language modeling, where the model is trained to predict the next token based on the previous context:
\begin{equation}
\label{eqn_1}
\begin{aligned}
L(C) = \mathbb{E}_{x \in C}\left\lbrack {\sum\limits_{x_{i} \in x}{\log{p\left( x_{i} \middle| x_{1},x_{2},\ldots,x_{i - 1} \right)}}} \right\rbrack
\end{aligned}
\end{equation}

{\setlength{\parindent}{0cm}where $C$ is a pre-training corpus and x is a sampled sequence from $C$. (2) De-noising auto-encoding, where the model is trained to restore a corrupted input sequence. In particular, masked language models (MLMs) \cite{KanekoMasahiro2020Encoder}, \cite{LabehatKryeziu2022ASoU} mask a subset of input tokens and predict them based on the remaining context:}

\begin{equation}
\label{eqn_1}
\begin{aligned}
L(C) = \mathbb{E}_{x \in C}\mathbb{E}_{\mathcal{M} \subset x,~{|\mathcal{M}|} = m{|x|}}\left\lbrack {\sum\limits_{x_{i} \in \mathcal{M}}{\log{p\left( x_{i} \middle| \overset{\sim}{x} \right)}}} \right\rbrack
\end{aligned}
\end{equation}

    {\setlength{\parindent}{0cm}where a mask $m$ (masking rate, typically 15\%) percentage of tokens from the original sentence $x$ and predicts the masked tokens $M$ given the corrupted context ~$x$ (the masked version of $x$).}

Different masking strategies have been proposed to sample $M$: Devlin et al. \cite{DevlinJacob2018BERT} randomly choose from the input tokens with a uniform distribution; Joshi et al. \cite{LiuYinhan2019RARO} sample contiguous spans of text; Levine et al. \cite{LevineYoav2020PPmo} sample words and spans with high Pointwise Mutual Information (PMI). These advanced sampling strategies prevent models from exploiting shallow local cues from uniform masking and lead to efficient pre-training. MLMs can encode bidirectional context while autoregressive language models can only “look at the past”, and thus MLMs are shown to be more effective at learning contextualized representations for downstream tasks \cite{DevlinJacob2018BERT}. On the other hand, MLMs suffer a significant computational cost because it only learns from 15\% of the tokens per sequence, whereas autoregressive LMs predict every token in a sequence. 

\subsection{AI4SE tasks}
Human life depends on reliable software; therefore, the software production process (i.e., software design \cite{Perez2022SD}, development \cite{Palomo-DuarteManuel2021Assessment}, and maintenance \cite{Maletic1994M}) becomes one of the most important factors to ensure the quality of software. With the increase in the complexity of software, how to improve the performance and efficiency of software production has become a challenge for software developers and researchers. To address this challenge, researchers have used information retrieval and DL technologies to implement a series of automated tools. These tools can solve SE tasks, such as code search, code summarization, and duplicate bug report detection.

Code search \cite{NiuHaoran2016Ltrc, GuXiaodong2018DeepCS, RaghothamanMukund2016SWIM, LiXuan2016Relationship-Aware, LvFei2015CodeHow} is frequently used by developers to conveniently find relevant code snippets. McMillan et al. \cite{McMillanCollin2011Pfrf} propose a code search engine that combines keyword matching with PageRank to return a chain of functions. Lv et al. \cite{LvFei2015CodeHow} propose CodeHow, a code search tool that incorporates an extended Boolean model and API matching. 
Ponzanelli et al. \cite{PonzanelliLuca2014Mstt} propose an approach that automatically retrieves pertinent discussions from Stack Overflow given a context in the integrated development environment (IDE). 
Code summarization automatically generates high-quality text to help developers understand the program. Sridhara et al. \cite{SridharaGiriprasad2010Tags} generate a code summary for the Java method from its method call and signature using the NLP techniques. Software Word Usage Model has been built for software analysis using NLP by Pollock et al. \cite{PollockLori2013NLSA}. 
Fowkes et al. \cite{FowkesJaroslav2017AfSC} design an unsupervised, extractive source code summarization system using an auto-folding method.
Programmers fix bugs based on bug reports. BM25F \cite{RobertsonStephen2004SBet} calculates the similarity between two bug reports based on common words shared between the bug reports. REP \cite{SunChengnian2011Tmar} extends BM25F by also considering bug report attribute information (e.g., product, priority). 
Deshmukh et al. \cite{DeshmukhJayati2017TADB} propose a deep learning-based approach (i.e., DLDBR), which mainly relies on the textual feature to detect duplicate bug reports. 
Our study is different from the above work. We study some tasks in program comprehension to demonstrate the overinterpretation in them.

More and more AI-based schemes have been proposed to solve traditional SE tasks \cite{AbbesM2011AESo,Perez2022SD,Palomo-DuarteManuel2021Assessment}. However, previous studies analyze the impact of different programming languages and antipatterns on program understanding \cite{PolitowskiCristiano2020Alse, AbbesM2011AESo, OlbrichS2009Teai}, as well as investigated the positive impact of dynamic analysis on program comprehension \cite{CornelissenB2009ASSo, NoughiNesrine2017AESo}. Moreover, researchers have studied sub-tasks of program understanding to help improve efficiency, such as code search \cite{YanShuhan2020AtCS}, code summarization \cite{McBurneyPaulW2014Aeso}, and duplicate bug report detection \cite{HaoRui2022Adrh}, etc.

\subsection{Overinterpretation}
Overinterpretation is a serious issue in black-box models. We define model overinterpretation to occur when a model finds strong class evidence in input that contains no semantically salient features. Overinterpretation is related to overfitting, but overfitting can be diagnosed via reduced test accuracy \cite{CarterBrandon2020Overinterpretation}. Overinterpretation may stem from real statistical signals in the distribution of the underlying data set that happen to be generated by specific properties of the data source. Thus, overinterpretation is harder to be diagnosed as it admits decisions that are made by statistically valid criteria, and models that use such criteria can excel at benchmarks. Meanwhile, PLMs always require a large dataset to train and fine-tune. Datasets always contain implicit biases and unique statistical signals. These biases and statistical signals often allow DL models to achieve high accuracy in test data by learning highly specific features unique to that dataset rather than generalizable features or key features under human understanding \cite{Chakraborty2022TSE}. However, these non-distinctive features are beyond human comprehension. They may be a series of unrelated characters or sparse pixels, and are not the features that we consider to be capable of making critical decisions. Although the outward appearance of this phenomenon can be surprising, as the model still works under bad conditions. But the model learns not problem-based features but dataset-based features. This can make the model less generalizable. 
Understanding overinterpretation is significant for improving model quality. Moreover, it can also provide guidance for designing the architecture of models.

\section{METHODOLOGY}
\label{sec_method}
\begin{figure*}{}
\centering
\includegraphics[width=0.8\textwidth]{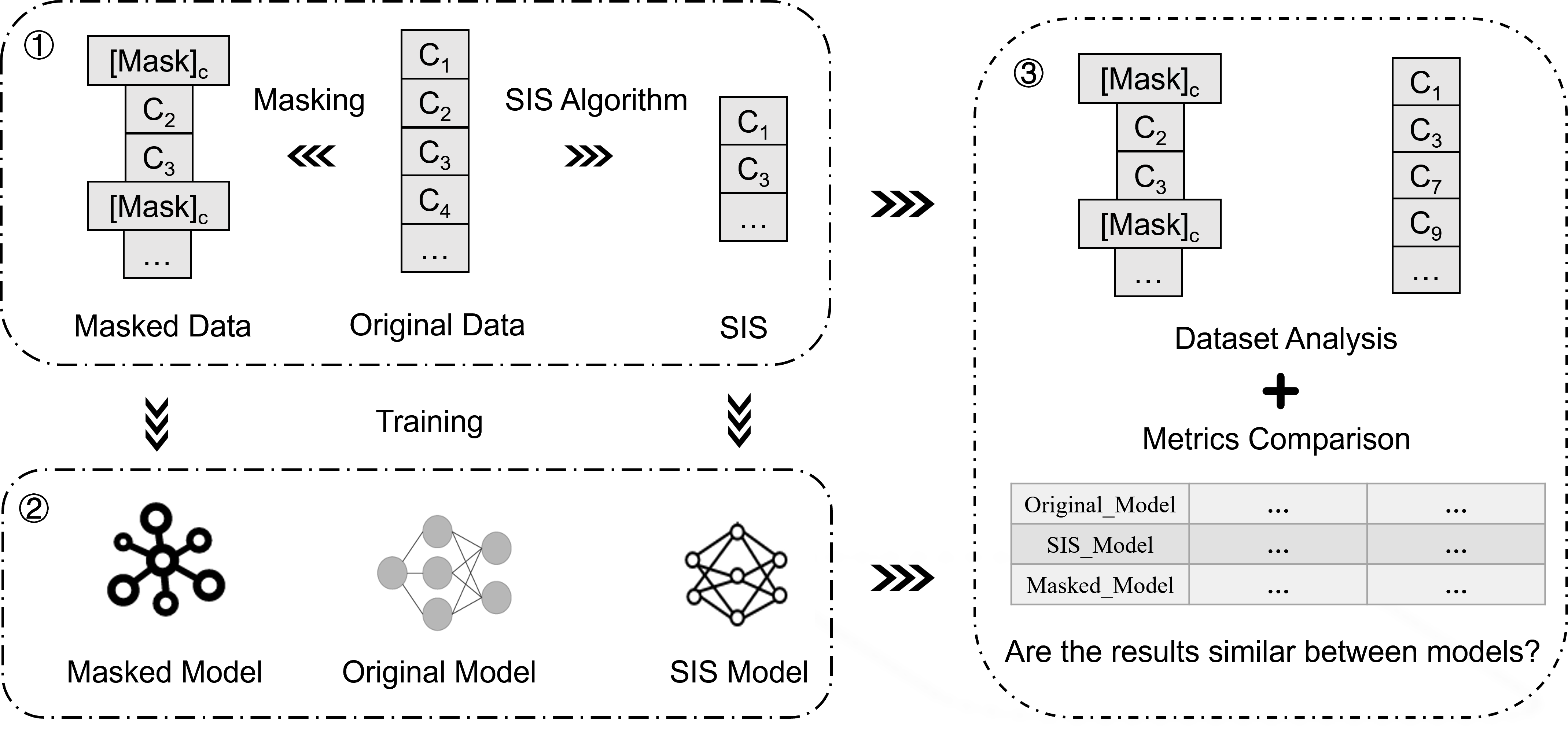}
\caption{\label{fig:overview}Flow of analysis in this study. Step \ding{172}, we process the raw data using the masking strategy and the SIS algorithm. Step \ding{173}, the models are trained using the processed data and the original data, respectively. Step \ding{174}, the three types of models and the features they use are compared and analyzed.}
\end{figure*}

\subsection{Overview}
Revealing overinterpretation requires a systematic way to identify which features are used by a model to reach its decision. In this study, to comprehensively examine whether pre-trained language models are overinterpreted, we propose two evaluation methods. One is to use multiple different masking rates to train the model. The other is to train the model using SIS. As shown in Fig. \ref{fig:overview}, our study consists of three main steps: 1) \textit{we use two methods to process the dataset. One is the masking rate strategy which masks the dataset with different degrees, and the other is to extract sufficient input subsets using the SIS algorithm;} 2) \textit{we train the PLM using the masked dataset and SIS separately;} 3) \textit{we use multiple evaluation metrics to comprehensively evaluate and compare the models, and analyze the corresponding datasets.}

\subsection{Different masking rate strategies}

The emergence of pre-trained models has made the training of large models easy. However, a masking strategy is introduced to perform sufficient representation learning during pre-training. It increases the learning difficulty of the model to some extent even though the model is more generalizable.

In this study, we set three masking rates, 15\% masking rate, 40\% masking rate, and 80\% masking rate. In addition to using the 15\% masking rate to train the model, we also use the 40\% and 80\% masking rates to train the model. A series of experiments are conducted to compare the differences between the classifiers trained with each masking rate. As the masking rate increases, the more corpus is masked, the more difficult it becomes for humans to understand its meaning. The performance of the PLM does not change much when the masking rate is increased from 15\% to 80\% (the evaluation metric stays within 10\% for both improvement and decrease). This fact indicates that the large reduction in input data has little impact on the effectiveness of the PLM. Meanwhile, as the masking rate increases, the utterance contains less information and becomes more difficult to understand. When the masking rate reaches 40\%, humans can no longer understand the meaning of the original sentence correctly. Thus, PLMs can learn information from unrelated or even meaningless words or letters to make the final decision. This fact proves that pre-trained language models overinterpret the data.

\subsection{Sufficient input subset}
The idea of SIS has been proposed to help humans interpret the decisions of black-box models \cite{CarterBrandon2018Black-box}. An SIS subset is a minimal subset of features that suffices to yield a class probability above a certain threshold with all other features masked. One simple explanation about why a particular black-box decision is reached may be obtained via a sparse subset of the input features whose values form the basis for the model’s decision – a rationale.

The SIS rationalizes why reaching a particular black-box decision only applies to input instances $x$ that satisfy the decision criterion $f(x)>m$. For such an input $x$, we aim to identify an SIS-collection of disjoint feature subsets that satisfy the following criteria:
\begin{itemize}
\item $f\left( X_{S_{n}} \right) \geq ~\tau~for~each~n~ = ~1,...~~,N$
\item There exists no feature subset $S^{'} \subset S_{n}$ for some $n=1,..., N$ such that $f\left( X_{S^{'}} \right) \geq \tau$
\item $f\left( X_{R} \right) < \tau~for~R~ = ~\lbrack p\rbrack~\backslash{\bigcup_{n = 1}^{N}S_{N}}$ (the remaining features outside of the SIS-collection)
\end{itemize}

Criterion (1) ensures that for any SIS ($S_n$), in the absence of any other features, the features in that subset alone are sufficient to justify the decision. Criterion (2) ensures that each SIS that reaches a decision contains a minimum number of features. Criterion (3) no longer reaches the same decision on the input after the entire SIS set is masked.

Thus, we perform the SIS algorithm to extract the subset of corpus and then train the corresponding model. By comparing the selected evaluation metrics, we consider the extracted SIS to be valid if the model trained using SIS is similar to the model trained using the full dataset. Then, we analyze the SIS to determine whether humans can understand the meaning of the SIS. If the SIS is not meaningful, it indicates that the model suffers from overinterpretation. 

\section{EXPERIMENTAL SETUP}
\label{sec_setup}

To ensure the authenticity of the experiments, all models use the default settings and the datasets are the same as those used in the original work. All the work investigated in the paper has public datasets and source code. The source code and dataset of code search tasks are provided by the previous studies \cite{GuXiaodong2018DeepCS,FengZhangyin2020CodeBERT}. We can find code summarization's source code and dataset in the literatures \cite{AhmadWasiUddin2020Transformer,IyerSrinivasan2016Summarizing}. Duplicate bug report detection task's code and dataset are in the following studies \cite{XiaoGuanping2020HINDBR,BudhirajaAmar2018DWEN}. The code snippets of PLMs can be found in the following studies \cite{Radford2018Corpus,DevlinJacob2018BERT,YangZhilin2019XLNET}. The evaluation indicators in this paper are the same as in the original work. All experiments are done on the same machine. We run our experiments on single NVIDIA Geforce 3090Ti GPU, Intel (R) Xeon(R) 2.60GHz 16 CPU. To comprehensively demonstrate the overinterpretation of the PLMs, we design two types of experiments. For the details, please refer to Section \ref{sec_task} and Section \ref{sec_model}. 

\section{TASK-ORIENTED OVERINTERPRETATION ANALYSIS}
\label{sec_task}

We study some AI4SE tasks and reveal the overinterpretation in these tasks. We choose three tasks, code search \cite{GuXiaodong2018DeepCS, FengZhangyin2020CodeBERT}, code summarization \cite{AhmadWasiUddin2020Transformer,IyerSrinivasan2016Summarizing}, and duplicate bug report detection \cite{XiaoGuanping2020HINDBR,BudhirajaAmar2018DWEN}. These tasks are not only widely used in software development and maintenance, but also encompass language processing techniques such as NL to PL translation, PL to NL translation, and NL classification. For each task, two representative studies are selected and their models are evaluated to detect the presence of overinterpretation. The next sections describe these experiments in detail.

\begin{tcolorbox}[title = {RQ 1:},
  colframe = gray!30!white, colback = gray!10!white,
  colbacktitle = gray!30!white,
  coltext = black!50!black,
  coltitle = black!90!white]
  Do software engineering tasks (code search, code summarization, and duplicate bug report detection) suffer from overinterpretation?
\end{tcolorbox}

\subsection{Code Search}
\label{sec_cs}

Code search is a frequent activity in software development. To implement a program functionality, developers can reuse previously written code snippets by searching through a large-scale codebase. Over the years, many code search tools have been proposed to help developers such as DeepCS \cite{GuXiaodong2018DeepCS} and CodeBERT \cite{FengZhangyin2020CodeBERT}, both of which use deep learning models to conduct code search. 

DeepCS is a novel code search tool using deep embedding neural networks. Instead of matching textual similarities, DeepCS co-embeds code snippets and natural language descriptions into a high-dimensional vector space, and then performs searches based on the vectors. They empirically evaluate DeepCS on a large-scale codebase collected from GitHub. The experimental results show that the method can effectively retrieve relevant code snippets and outperforms previous techniques. CodeBERT is a bimodal pre-trained model for a programming language and natural language. Authors evaluate CodeBERT on two NL-PL applications by fine-tuning model parameters. Results show that CodeBERT achieves state-of-the-art performance on both natural language code search and code documentation generation.

\begin{table}[]
\centering
\caption{Overall Accuracy of DeepCS under different masking rates. ``MR" means masking rate. ``NA" means that no masking strategy is used.}
\label{mask_c1}
\begin{tabular}{cccccccc}
\hline
Pre-training & \multicolumn{7}{c}{Metrics}                    \\ \hline
MR   & R@1  & R@5  & R@10 & P@1  & P@5  & P@10 & MRR  \\ \hline
NA           & 0.51 & 0.82 & 0.89 & 0.49 & 0.51 & 0.48 & 0.61 \\ \hline
15\%         & 0.48 & 0.76 & 0.85 & 0.48 & 0.47 & 0.46 & 0.60 \\ \hline
40\%         & 0.47 & 0.75 & 0.82 & 0.46 & 0.44 & 0.45 & 0.59 \\ \hline
80\%         & 0.42 & 0.69 & 0.77 & 0.41 & 0.42 & 0.43 & 0.53 \\ \hline
\end{tabular}

\end{table}

\begin{table}[]
\centering
\caption{Overall Accuracy of CodeBERT under different masking rates. ``MR" means masking rate.}
\label{mask_c2}
\begin{tabular}{ccccccc}
\hline
MR & RUBY & JAVASCRIPT & GO   & PYTHON & JAVA & PHP  \\ \hline
NA         & 0.69 & 0.71       & 0.84 & 0.86   & 0.74 & 0.70 \\ \hline
15\%       & 0.71 & 0.72       & 0.83 & 0.87   & 0.74 & 0.72 \\ \hline
40\%       & 0.74 & 0.70       & 0.85 & 0.88   & 0.75 & 0.71 \\ \hline
80\%       & 0.67 & 0.68       & 0.80 & 0.82   & 0.71 & 0.69 \\ \hline
\end{tabular}

\end{table}

First, we train models using different masking rates, 15\% masking rate, 40\% masking rate, and 80\% masking rate. Four common metrics are used to measure the effectiveness of code search, namely, FRank \cite{RaghothamanMukund2016SWIM}, SuccessRate@k \cite{LiXuan2016Relationship-Aware}, Precision@k \cite{LvFei2015CodeHow}, and Mean Reciprocal Rank (MRR) \cite{SafdariNasir2018LtRR}. They are widely used metrics in information retrieval and code search literature. Table \ref{mask_c1} shows the performance of DeepCS under different masking rates. DeepCS outperforms other models when training models using the full dataset. But after using the masking rate strategy to train the model, the performance of DeepCS decreases. Moreover, as the masking rate increases, the performance of DeepCS gradually decreases. When the model is trained with 80\% masking rate, DeepCS has the lowest performance, but the difference is within 10\% compared to the model trained with the complete dataset. This fact demonstrates that DeepCS can learn enough ``knowledge" to make accurate judgments with most of the input missing (at least 20\% of the data is retained). ``Knowledge" refers to the real statistical signals in the dataset. Meanwhile, the high masking rate means that it is difficult for humans to understand masked sentences. We conduct the same experiment with CodeBERT. Table \ref{mask_c2} shows the performance of CodeBERT. Unlike DeepCS, the performance of CodeBERT does not decrease as the masking rate increases, but increases. This result indicates that the performance of CodeBERT improves as the input dataset is reduced. Moreover, CodeBERT can still make accurate judgments when the masking rate is 80\%, i.e., only 20\% of the input data is retained. This suggests that the above models do not learn the features in the dataset but the statistical signals in the dataset. 

\begin{figure}[ht]
    \centering
    \includegraphics[scale=0.38]{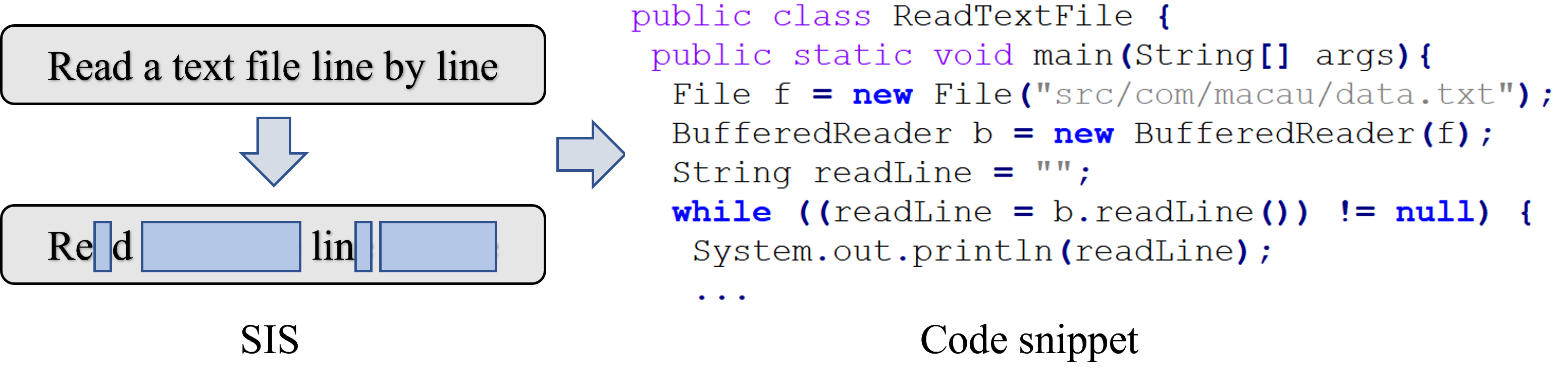}
    \caption{Example of SIS in the code search. The blue blocks mean the data filtered out by the SIS algorithm.}
    \label{fig:cs_SIS}

\end{figure}

\begin{figure*}[ht]
\centering
\subfigure[Performance of DeepCS under SIS]{%
\includegraphics[align=c,scale=0.37]{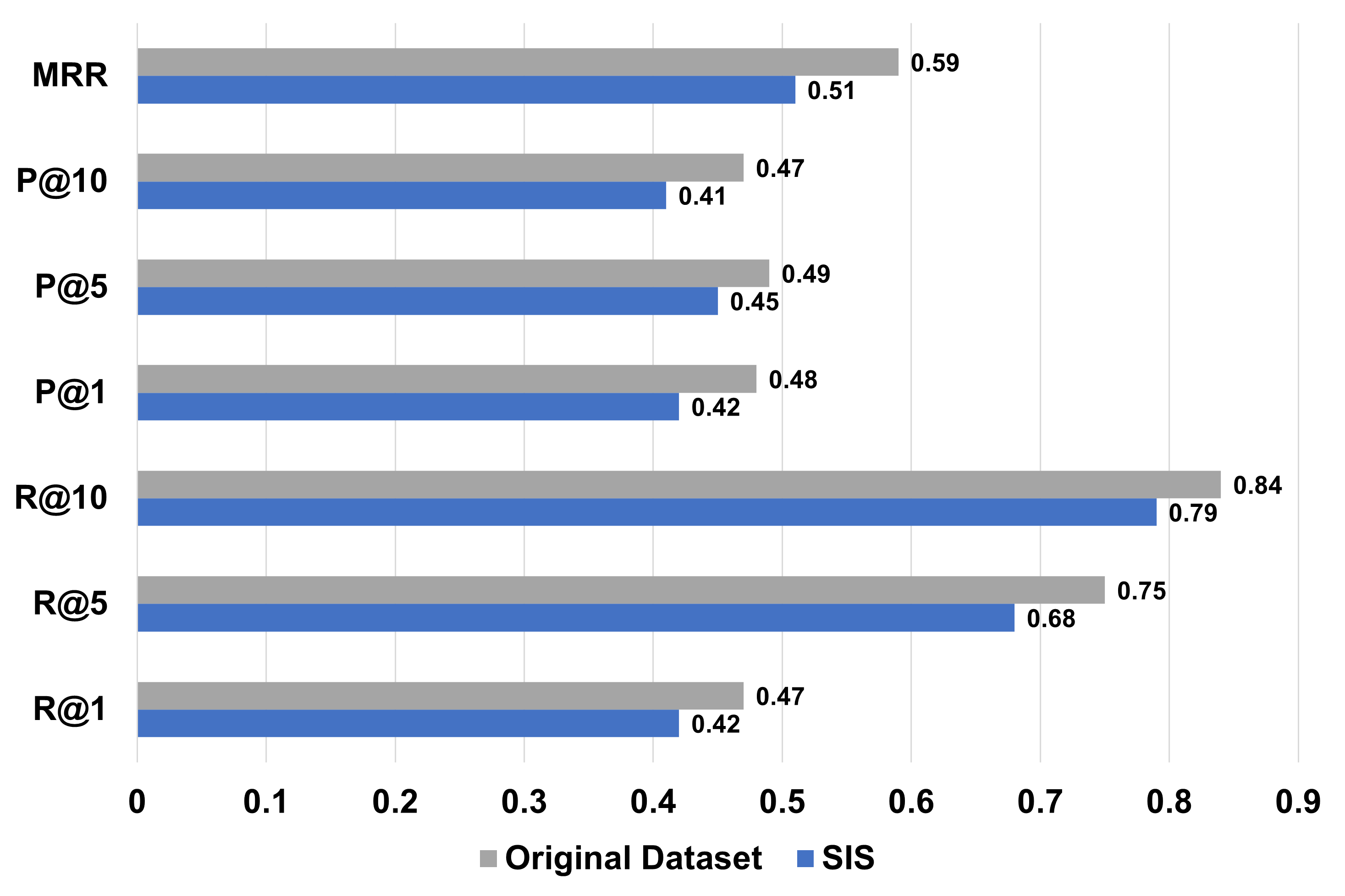}
\label{fig:cs_sis1}}
\quad
\subfigure[Performance of CodeBERT under SIS]{%
\includegraphics[align=c,scale=0.37]{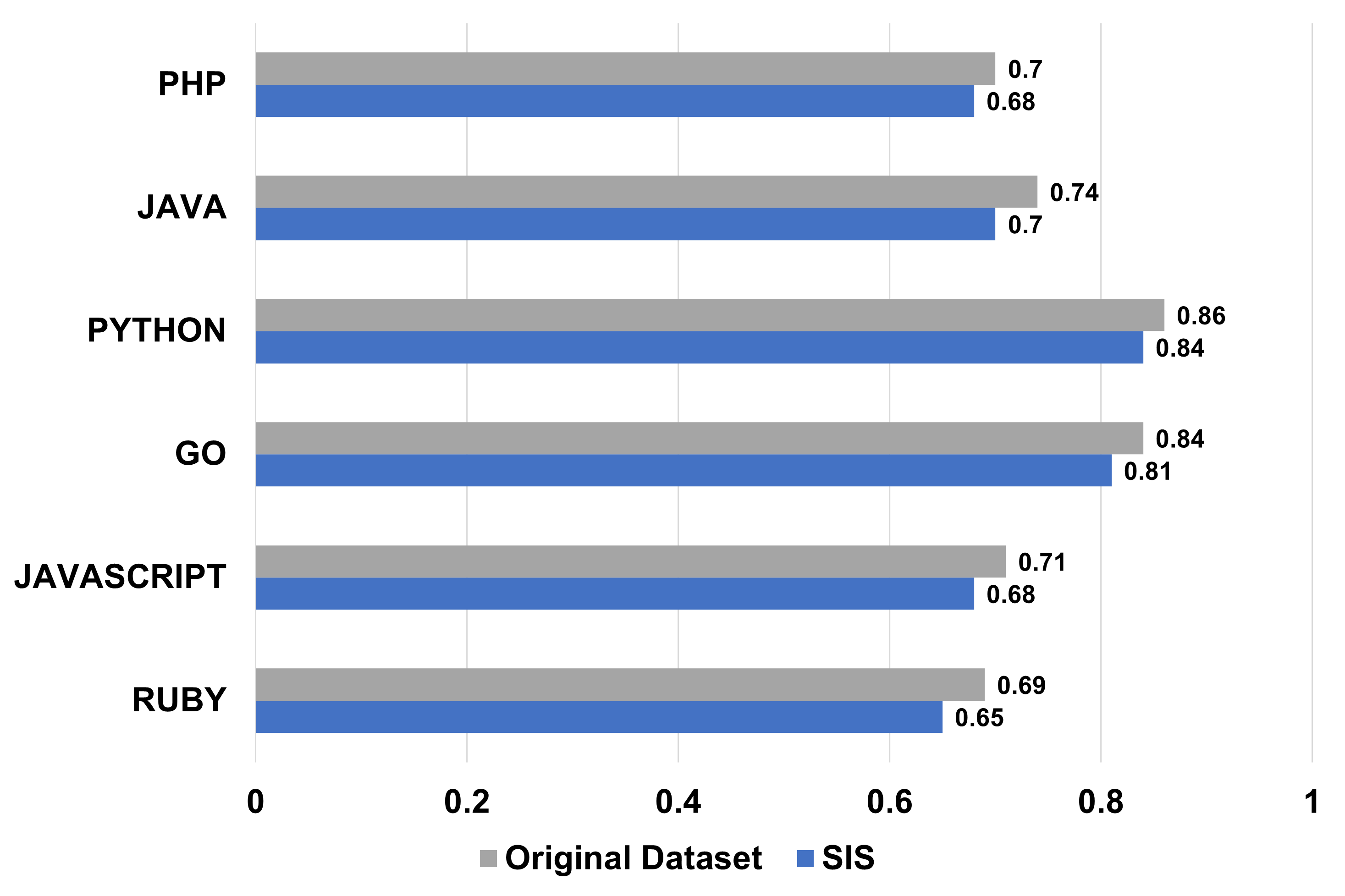}
\label{fig:cs_sis2}}
\centering
\caption{The performance of SIS in the code search.}
\label{fig:cs_sis_d}

\end{figure*}

Then, we use the SIS algorithm to extract a subset to train DeepCS, and CodeBERT. Results are shown in Fig. \ref{fig:cs_sis1} and Fig. \ref{fig:cs_sis2}. SIS is a subset of the complete dataset, which is sparse but allows the model to make accurate decisions. To demonstrate SIS in more detail, we show an example of code search, as shown in Fig. \ref{fig:cs_SIS}. The original query is ``Read a text file line by line", after using the SIS algorithm, the extracted SIS is shown below. SIS contains ``red", and ``lin". SIS is completely different from the original query. For humans, the salient features of the query are ``read" and ``line", and based on these two words, humans can make some inferences to search code snippet. But now we cannot understand what the SIS represents, much less find the corresponding code snippet based on the SIS. However, for a pre-trained language model, it can still search for the corresponding code fragment based on the subset. The results are shown in Fig. \ref{fig:cs_sis_d}. While retaining very little input data, the model achieves good results. The results are similar to the model trained with the full dataset. This suggests that pre-trained language models do not make decisions based on salient features, but learn statistical signals unique to the dataset to make judgments. In the absence of salient features, DeepCS and CodeBERT can still achieve good performance. Therefore, the tasks related to code search overinterpret the dataset.

\begin{tcolorbox}[
  colframe = gray!30!white, colback = gray!10!white,
  colbacktitle = gray!30!white,
  coltext = black!50!black,
  coltitle = black!90!white]
{\setlength{\parindent}{0cm}
\textbf{Finding 1.1:} Code search suffers from overinterpretation. PLMs trained in different ways have similar performance, and these models are trained without salient features. } 
\end{tcolorbox}

\subsection{Code Summarization}
\label{sec_summ}

Generating a readable summary that describes the functionality of a program is known as source code summarization which can help developers understand and maintain software. In this task, learning code representation by modeling the pairwise relationship between code tokens to capture their long-range dependencies is crucial. To learn code representation for summarization, researchers have proposed many efficient methods such as \cite{AhmadWasiUddin2020Transformer} and \cite{IyerSrinivasan2016Summarizing}. 

Ahmad et al. \cite{AhmadWasiUddin2020Transformer} explore the Transformer model that uses a self-attention mechanism and has shown to be effective in capturing long-range dependencies. The authors perform experiments on two well-studied datasets, and the results endorse the effectiveness.

Iyer et al. \cite{IyerSrinivasan2016Summarizing} present, CODE-NN, the first completely data-driven approach for generating high-level summaries of source code. Experiments outperform strong baselines.

\begin{table}[]
\centering
\caption{Overall performance of code summarization tasks under different masking rates. ``MR" means masking rate.}
\label{mask_summ}
\begin{tabular}{ccccc}
\hline
\multirow{2}{*}{Works}             & Pre-T & \multicolumn{3}{c}{Metrics} \\ \cline{2-5} 
                                   & MR   & BLEU   & METEOR  & ROUGE-L  \\ \hline
\multirow{4}{*}{Transformer-based} & NA           & 0.44   & 0.26    & 0.54     \\ \cline{2-5} 
                                   & 15\%         & 0.45   & 0.25    & 0.55     \\ \cline{2-5} 
                                   & 40\%         & 0.47   & 0.32    & 0.61     \\ \cline{2-5} 
                                   & 80\%         & 0.46   & 0.28    & 0.57     \\ \hline
\multirow{4}{*}{CODE-NN}           & NA           & 0.25   & 0.17    & 0.56     \\ \cline{2-5} 
                                   & 15\%         & 0.22   & 0.15    & 0.53    \\ \cline{2-5} 
                                   & 40\%         & 0.21   & 0.13    & 0.51     \\ \cline{2-5} 
                                   & 80\%         & 0.18   & 0.10    & 0.48     \\ \hline
\end{tabular}

\end{table}

\begin{figure}[ht]
    \centering
    \includegraphics[scale=0.38]{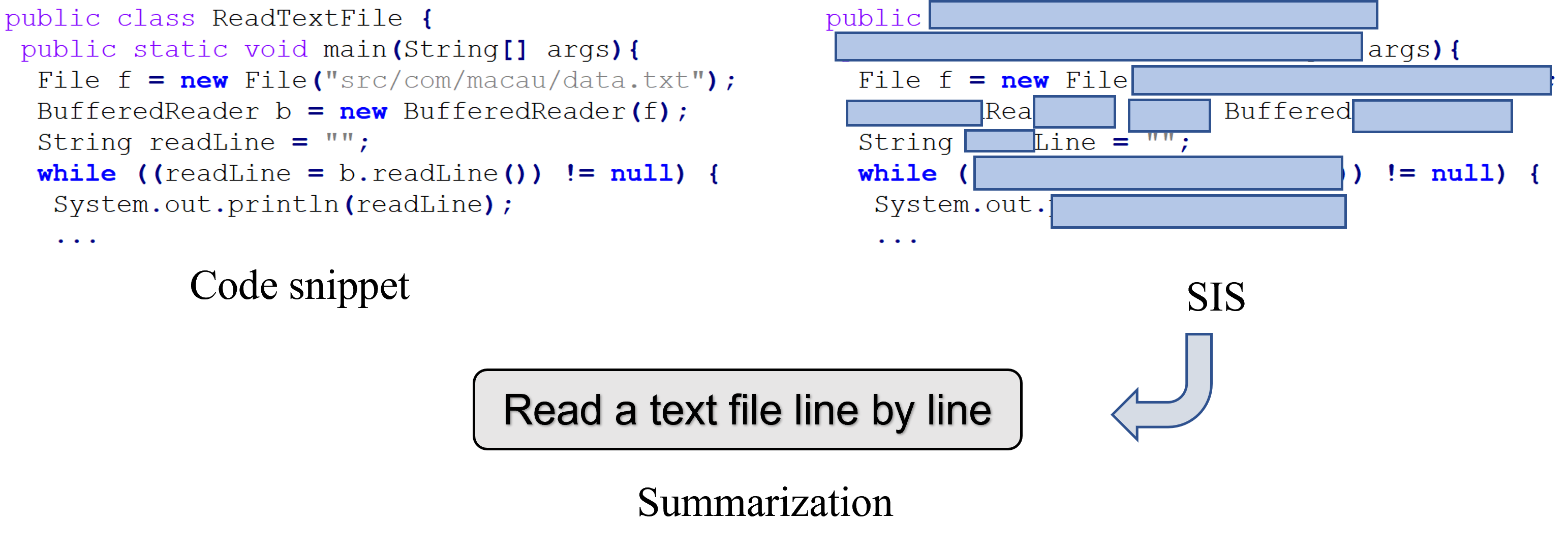}
    \caption{Example of SIS in the code summarization. The blue blocks mean the data filtered out by the SIS algorithm.}
    \label{fig:summ-e}

\end{figure}

\begin{figure*}[ht]
\centering
\subfigure[The performance of Transformer-based using SIS.]{%
\includegraphics[align=c,scale=0.37]{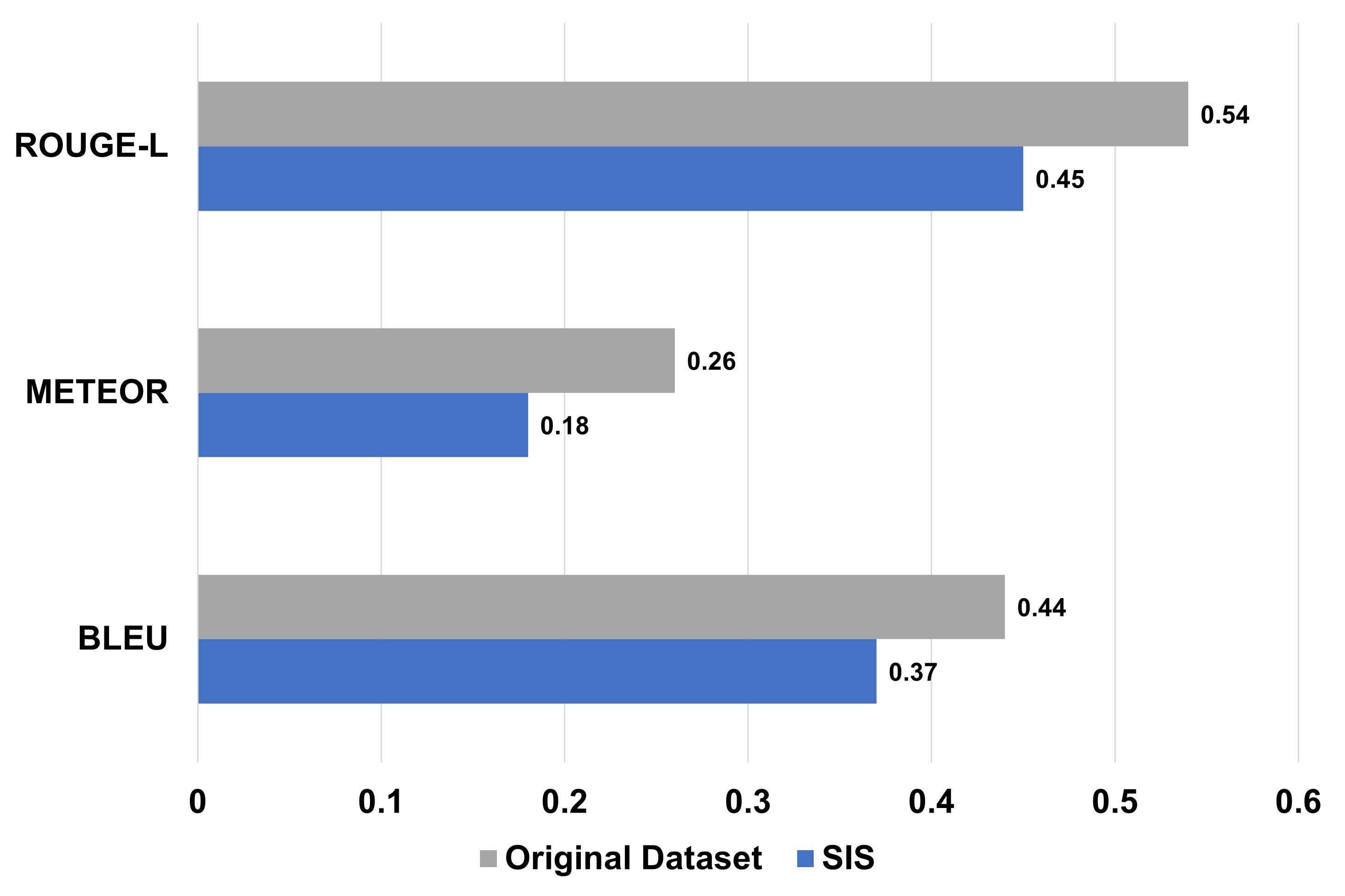}
\label{fig:summ1}}
\quad
\subfigure[The performance of CODE-NN under SIS.]{%
\includegraphics[align=c,scale=0.37]{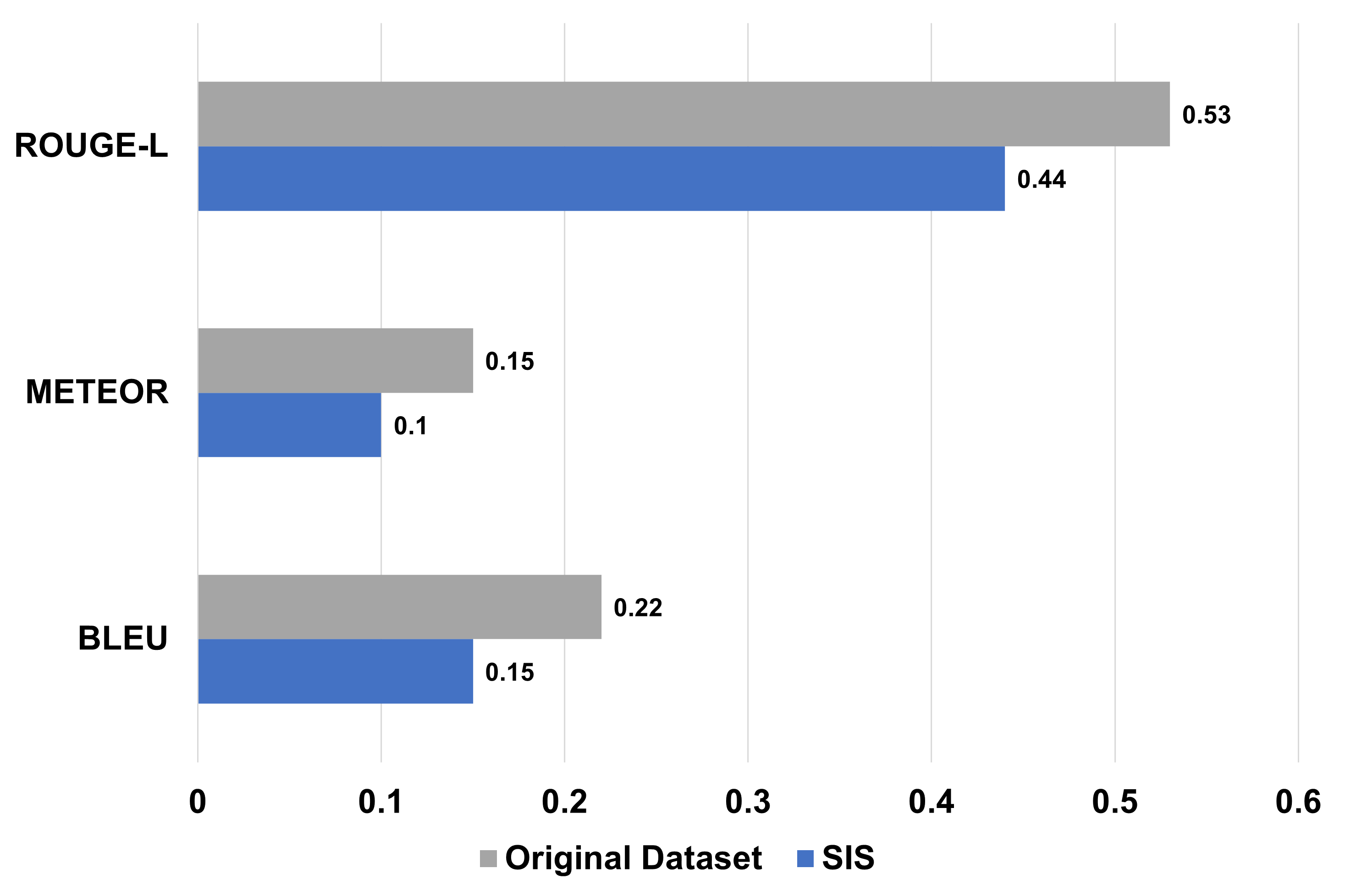}
\label{fig:summ2}}
\caption{The performance of code summarization under SIS.}
\label{fig:summ}

\end{figure*}

We choose the above research studies to investigate whether there is overinterpretation. The experimental results are shown in Table \ref{mask_summ}. We evaluate the source code summarization performance using three metrics, BLEU \cite{PapineniKishore2002BLEU}, METEOR \cite{BanerjeeSatanjeev2005Meteor}, and ROUGE-L \cite{lin2004rouge}. The performance of Transformer-based shows a pattern of increasing and then decreasing as the masking rate increases. In addition, the 80\% masking rate outperforms the 15\% masking rate. This result shows that pre-trained language models can make accurate decisions based on either the complete dataset or only 20\% of the dataset retained. It also proves that not all data are useful for Transformer-based. CODE-NN's performance decreases as the masking rate increases. Although each metric decreases, the difference in metrics is small (within 10\%) compared to the model trained on the full dataset. CODE-NN learns approximately for the dataset, regardless of the amount of data contained in the dataset. With the loss of a large amount of data, the CODE-NN can still make accurate judgments, and the only remaining datasets are completely incomprehensible to humans. Fig. \ref{fig:summ-e} shows an example of training code summarization using the SIS. 
The left part shows the original code snippet, and the right part shows the subset extracted by the SIS algorithm. Note that the text is generated based on this subset. A code fragment contains many salient features, such as class names, function names, etc. However, in the masked code snippet, the function names and class names that can indicate the purpose of the code are masked. Humans cannot understand the function of this code snippet by reading the masked code. But PLMs still generate accurate code descriptions. Fig. \ref{fig:summ1} and Fig. \ref{fig:summ2} describe the performance of the model after training with SIS. Although the performance of both tasks decreased after training with SIS. But the difference with the model trained using the full dataset is small. The code summarization model can still achieve good results in the absence of salient features. Therefore, there is an overinterpretation of the code summarization task.

\begin{tcolorbox}[
  colframe = gray!30!white, colback = gray!10!white,
  colbacktitle = gray!30!white,
  coltext = black!50!black,
  coltitle = black!90!white]
{\setlength{\parindent}{0cm}
\textbf{Finding 1.2:} The model is able to generate accurate summaries despite the absence of salient features. The experimental results show that the model overinterpret the dataset.}
\end{tcolorbox}

\subsection{Duplicate bug report detection}
\label{sec_dbr}

Bug report filing is a major part of software maintenance. Developers rely on bug reports to fix bugs. Due to different expression habits, different reporters may use different expressions to describe the same bug in the bug tracking system. As a result, the bug tracking system usually contains many duplicate bug reports. Automated duplicate detection can reduce developers' workload on fixing duplicate bugs. In other words, capturing and tagging duplicate bug reports is crucial to avoid the assignment of the same bug to different developers. Efforts have been made in the past to detect duplicate bug reports by using deep learning methods \cite{XiaoGuanping2020HINDBR}, \cite{BudhirajaAmar2018DWEN}. 

Xiao et al. \cite{XiaoGuanping2020HINDBR} present HINDBR, a novel deep neural network (DNN) that accurately detects semantically similar duplicate bug reports using a heterogeneous information network (HIN). Results show that HINDBR is effective. Budhiraja et al. \cite{BudhirajaAmar2018DWEN} propose Deep Word Embedding Network (DWEN) that uses a deep word embedding network for duplicate bug report detection. DWEN computes the similarity between two bug reports for duplicate bug report detection. Results show that the proposed approach is able to perform better than baselines.

\begin{table}[]
\caption{Overall performance of duplicate bug report detection tasks under different masking rates. ``MR" means masking rate. ``Pre-T" represents pre-trained.}
\label{mask_d}
\begin{tabular}{cccccc}
\hline
\multirow{2}{*}{Works}  & Pre-T & \multicolumn{4}{c}{Metrics}              \\ \cline{2-6} 
                        & MR   & Accuracy & Precision & Recall & F1 Score \\ \hline
\multirow{4}{*}{HINDBR} & NA           & 0.96     & 0.91      & 0.88   & 0.87    \\ \cline{2-6} 
                        & 15\%         & 0.94     & 0.89      & 0.83   & 0.86     \\ \cline{2-6} 
                        & 40\%         & 0.92     & 0.86      & 0.80   & 0.84      \\ \cline{2-6} 
                        & 80\%         & 0.89     & 0.81      & 0.79   & 0.78     \\ \hline
\multirow{4}{*}{DWEN}   & NA           & 0.82     & 0.73      & 0.79   & 0.78     \\ \cline{2-6} 
                        & 15\%         & 0.80     & 0.70      & 0.76   & 0.77     \\ \cline{2-6} 
                        & 40\%         & 0.75     & 0.67      & 0.74   & 0.73     \\ \cline{2-6} 
                        & 80\%         & 0.74     & 0.65      & 0.70   & 0.68     \\ \hline
\end{tabular}

\end{table}

\begin{figure}[ht]
    \centering
    \includegraphics[scale=0.4]{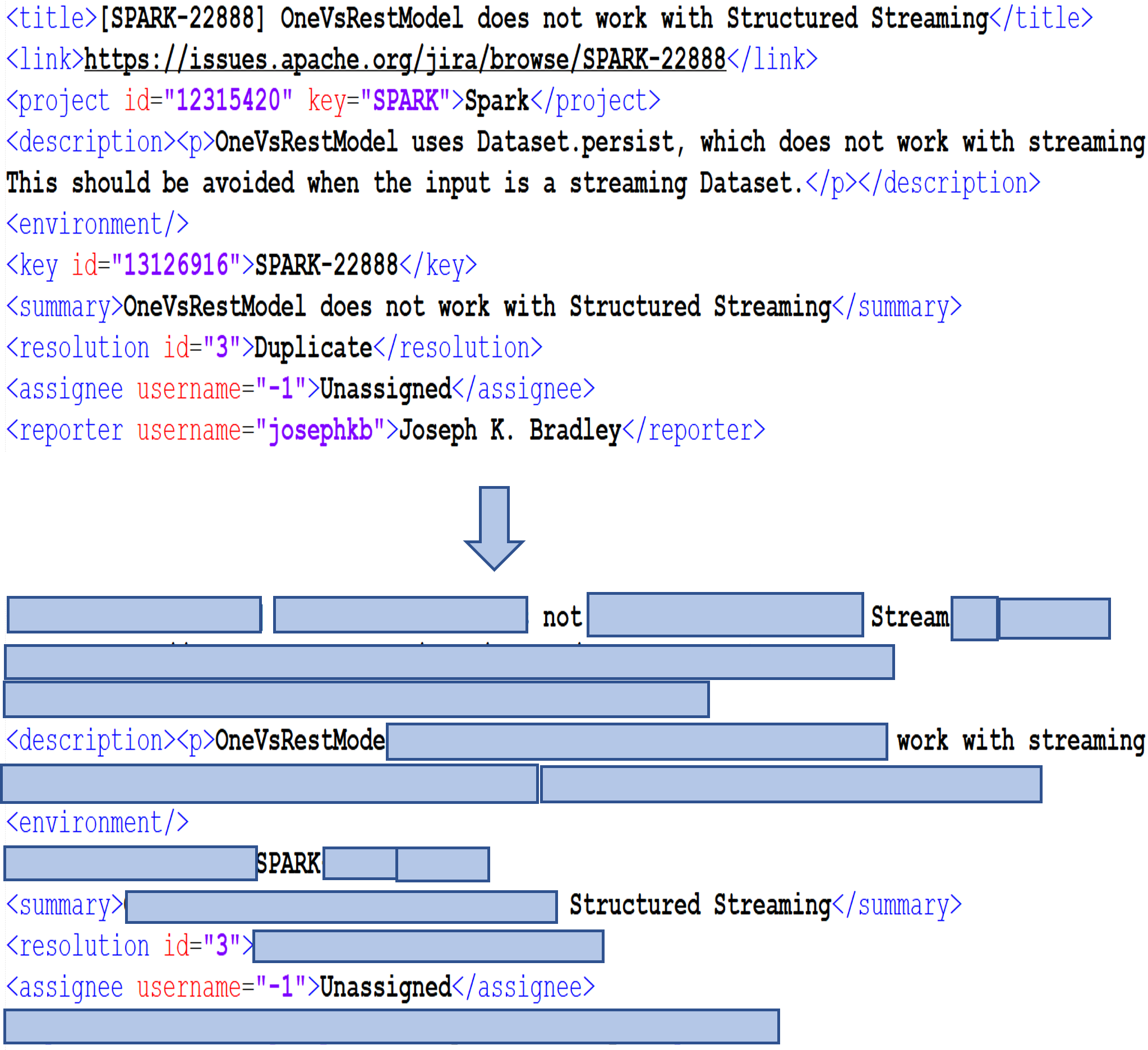}
    \caption{Example of SIS in the duplicate bug report detection tasks. The blue blocks mean the data filtered out by the SIS algorithm. }
    \label{fig:dbrd}
    \vspace{-0.5cm}
\end{figure}

\begin{figure*}[ht]
\centering
\subfigure[Performance of HINDBR under SIS.]{%
\includegraphics[align=c,scale=0.37]{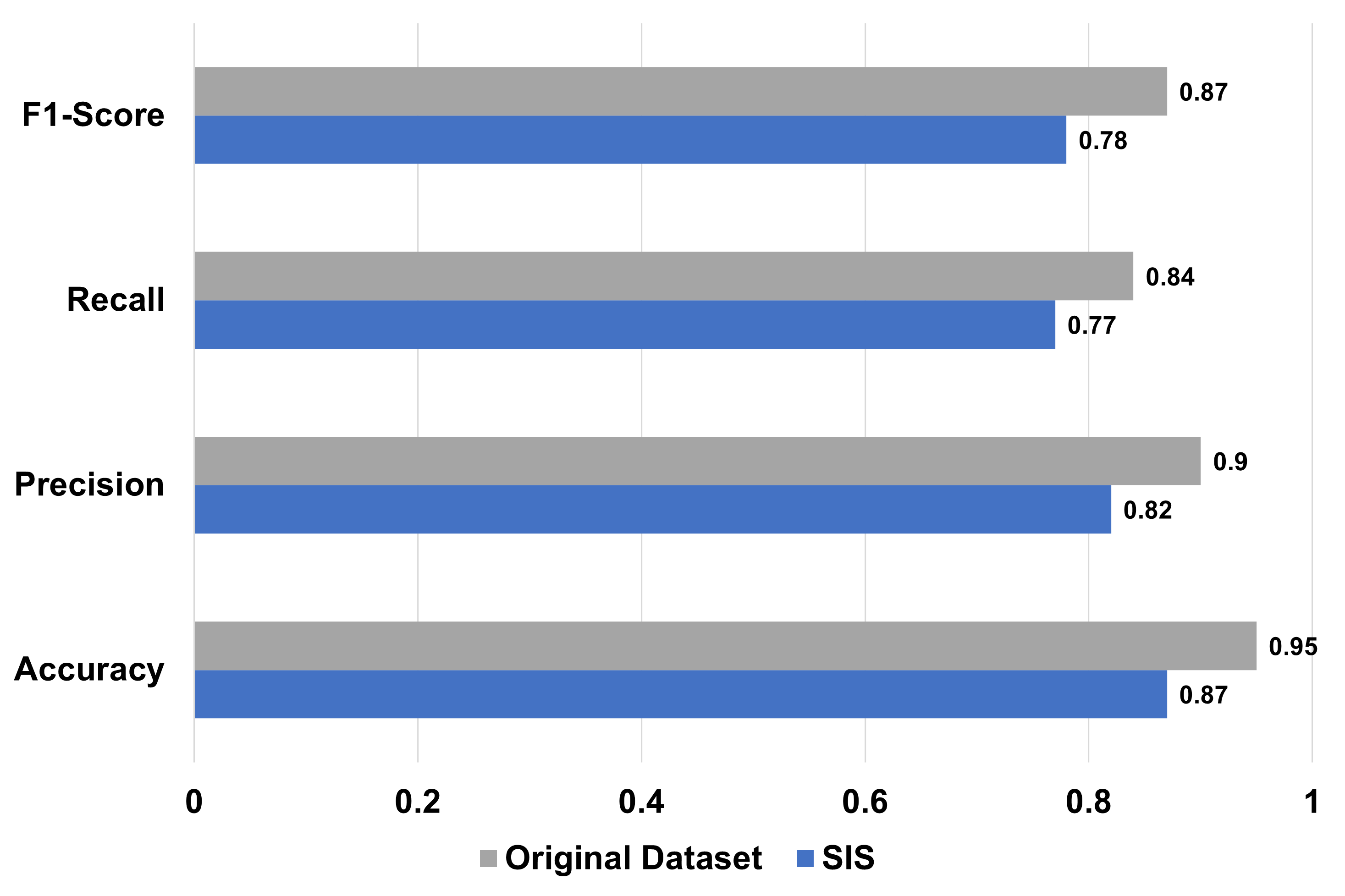}
\label{fig:dbrd1}}
\quad
\subfigure[Performance of DWEN under SIS.]{%
\includegraphics[align=c,scale=0.37]{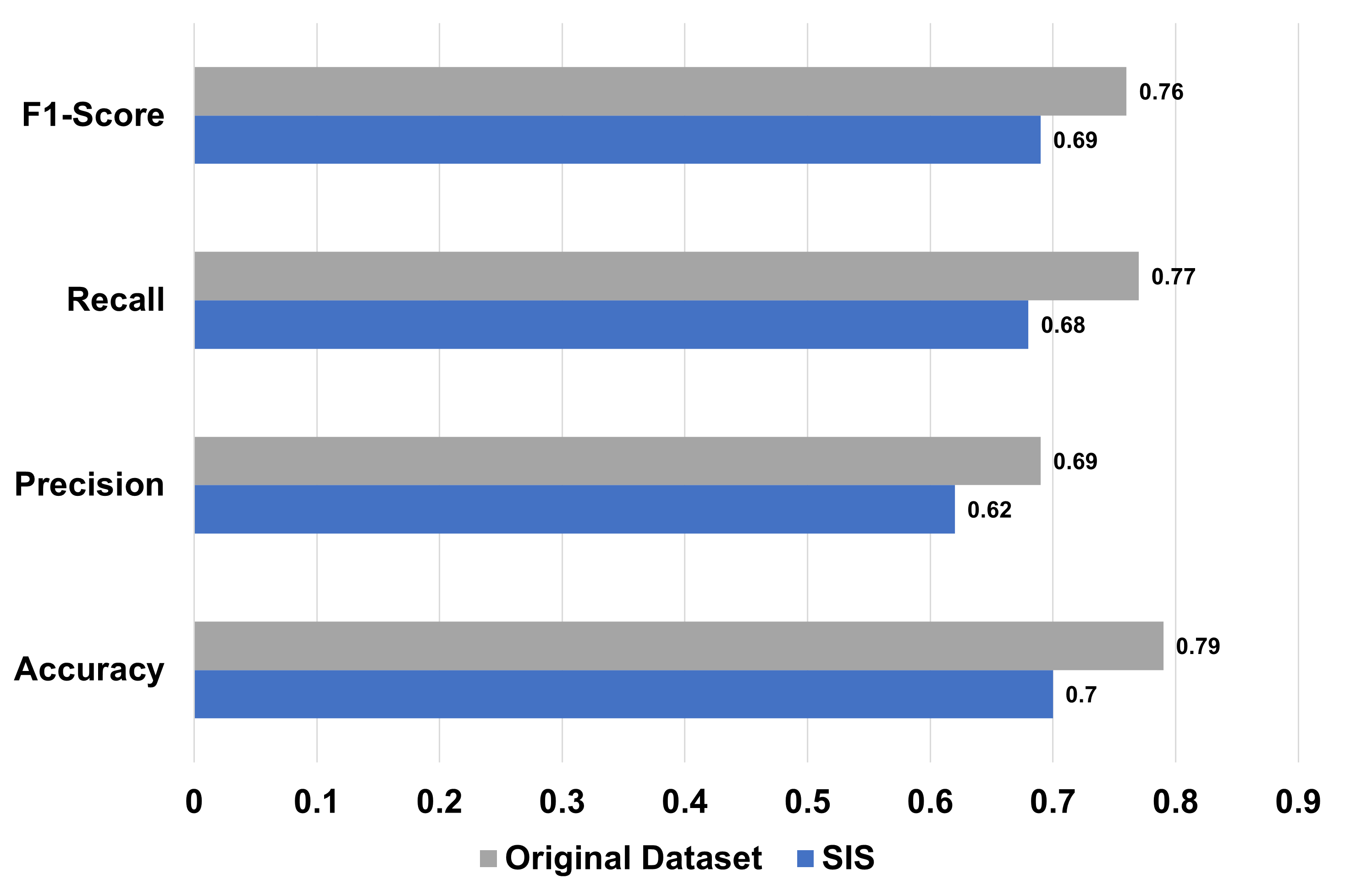}
\label{fig:dbrd2}}
\caption{The performance of duplicate bug report detection tasks under SIS.}
\label{fig:cs_dbrd}
\end{figure*}

Table \ref{mask_d} shows the metrics of both works under different masking rates, 15\% masking rate, 40\% masking rate, and 80\% masking rate. To evaluate models, we use the following four metrics, Accuracy, Precision, Recall, and F1-Score. When the model is trained using the masking rate strategy, the performance of the model starts to decrease. As the masking rate increases, all the metrics of the model decrease. The effect of the model is minimized when the masking rate reaches 80\%. However, the difference between all metrics and the original model is maintained within 10\%. This fact demonstrates that HINDBR and DWEN cannot learn key representations from the rich dataset. Fig. \ref{fig:dbrd} displays an example of using SIS to detect duplicate bug reports. The top part shows the complete bug report, and the bottom part shows the bug report after masking. The key to detecting duplicate bug reports is to learn the characterization of the report based on its description. However, the content in the description tag is mostly obscured. The remaining content in the description tag is meaningless to humans. Moreover, the results of the two schemes using SIS detection are depicted in Fig. \ref{fig:cs_dbrd}. Although all four metrics decline for the model trained with SIS, the difference with the original model (the model trained with the full dataset) stays within 10\%. Thus, there is an overinterpretation phenomenon in the studies related to duplicate bug report detection.

\begin{tcolorbox}[
  colframe = gray!30!white, colback = gray!10!white,
  colbacktitle = gray!30!white,
  coltext = black!50!black,
  coltitle = black!90!white]
{\setlength{\parindent}{0cm}
\textbf{Finding 1.3:} Lacking a salient part of the description, the model can still find duplicate reports. The PLMs in the duplicate bug report detection task suffers from overinterpretation.}
\end{tcolorbox}

\subsection{Analysis of SE tasks}
\label{sec_pc}

To understand whether the phenomenon of overinterpretation is prevalent in AI4SE tasks, we choose three types of tasks for our experiments, code search \cite{GuXiaodong2018DeepCS, FengZhangyin2020CodeBERT}, code summarization \cite{AhmadWasiUddin2020Transformer,IyerSrinivasan2016Summarizing}, and duplicate bug report detection \cite{XiaoGuanping2020HINDBR,BudhirajaAmar2018DWEN}. Tables \ref{mask_c1}-\ref{mask_d} present the results under different masking rates, 15\% masking rate, 40\% masking rate, and 80\% masking rate. First, in the software development phase, we analyze the code search task. We use a variety of approaches to train DeepCS and CodeBERT. Moreover, we present detailed examples to describe the final data used for model decisions. The analysis shows that the selected studies can find the right code snippets in the absence of prominent representations. Second, code summarization is the foundation of software maintenance. The Transformer-based and CODE-NN can generate an accurate summary based on the masked code snippet. Finally, in the software maintenance phase, we compare two duplicate bug report detection approaches. HINDBR and DWEN can accurately detect duplicate bug reports after masking the content of the description tag. Meanwhile, to more comprehensively assess whether overinterpretation exists in the AI4SE task, we also present examples for a more detailed description, as shown in Figures \ref{fig:cs_SIS}, \ref{fig:summ-e}, and \ref{fig:dbrd}. Finally, we show the performance of different models after training with SIS in Figures \ref{fig:cs_sis_d}, \ref{fig:summ}, and \ref{fig:cs_dbrd}. These experimental results all show that PLMs in SE tasks achieve high performance despite the lack of salient features. However, these inputs without salient features, are meaningless. It contains only discrete letters and sparse words. Humans simply cannot read the masked text, let alone understand its meaning. These models do not really understand these SE tasks, but only learn meaningless statistical signals. This is an overinterpretation phenomenon of the SE tasks.

\section{MODEL-ORIENTED OVERINTERPRETATION ANALYSIS}
\label{sec_model}

\begin{tcolorbox}[title = {RQ 2:},
  colframe = gray!30!white, colback = gray!10!white,
  colbacktitle = gray!30!white,
  coltext = black!50!black,
  coltitle = black!90!white]
  Does overinterpretation depend on software engineering tasks and how prevalent is overinterpretation in PLMs?  
\end{tcolorbox}

For various tasks of SE, researchers not only construct their models but also use well-known PLMs \cite{Radford2018Corpus,DevlinJacob2018BERT,YangZhilin2019XLNET}. Pre-trained models are beneficial for downstream NLP tasks and can avoid training a new model from scratch. To demonstrate that overinterpretation is not task-dependent but is prevalent in pre-trained language models, we choose three representative PLMs, GPT \cite{Radford2018Corpus}, BERT \cite{DevlinJacob2018BERT}, and XLNet \cite{YangZhilin2019XLNET} for evaluation. We perform the same training strategy for each model, i.e., different masking rates (15\% masking rate, 40\% masking rate, and 80\% masking rate) and SIS. We evaluate them with diverse tasks, including , multi-genre natural language inference corpus (MNLI) \cite{NangiaNikita2017TR2S}, question-answering natural language inference (QNLI) \cite{RajpurkarPranav2016SQuAD}, recognizing textual entailment (RTE), corpus of linguistic acceptability (CoLA) \cite{Warstadt2018Corpus}, the Stanford Sentiment Treebank (SST-2) \cite{SocherRichard2013Recursive}, the Microsoft Paraphrase corpus (MRPC) \cite{Dolan2022Automatically}, the Quora Question Pairs (QQP) \cite{Chen2018Quora}, and the Semantic Textual Similarity benchmark (STS-B) \cite{CerD2017ST1s}. These downstream tasks are widely used to evaluate PLMs.

\subsection{GPT}
The GPT \cite{Radford2018Corpus} family is a series of very powerful pre-trained language models proposed by OpenAI, which can achieve stunning results in very complex tasks, such as article generation, code generation, machine translation, etc., without the need for supervised learning for model fine-tuning. In contrast to previous approaches, they make use of task-aware input transformations during fine-tuning to achieve effective transfer while requiring minimal changes to the model architecture. For a new task, GPT requires very little data to understand the requirements of the task and to approach or exceed the state-of-the-art approach.

\begin{figure*}[ht]
    \centering
    \includegraphics[scale=0.7]{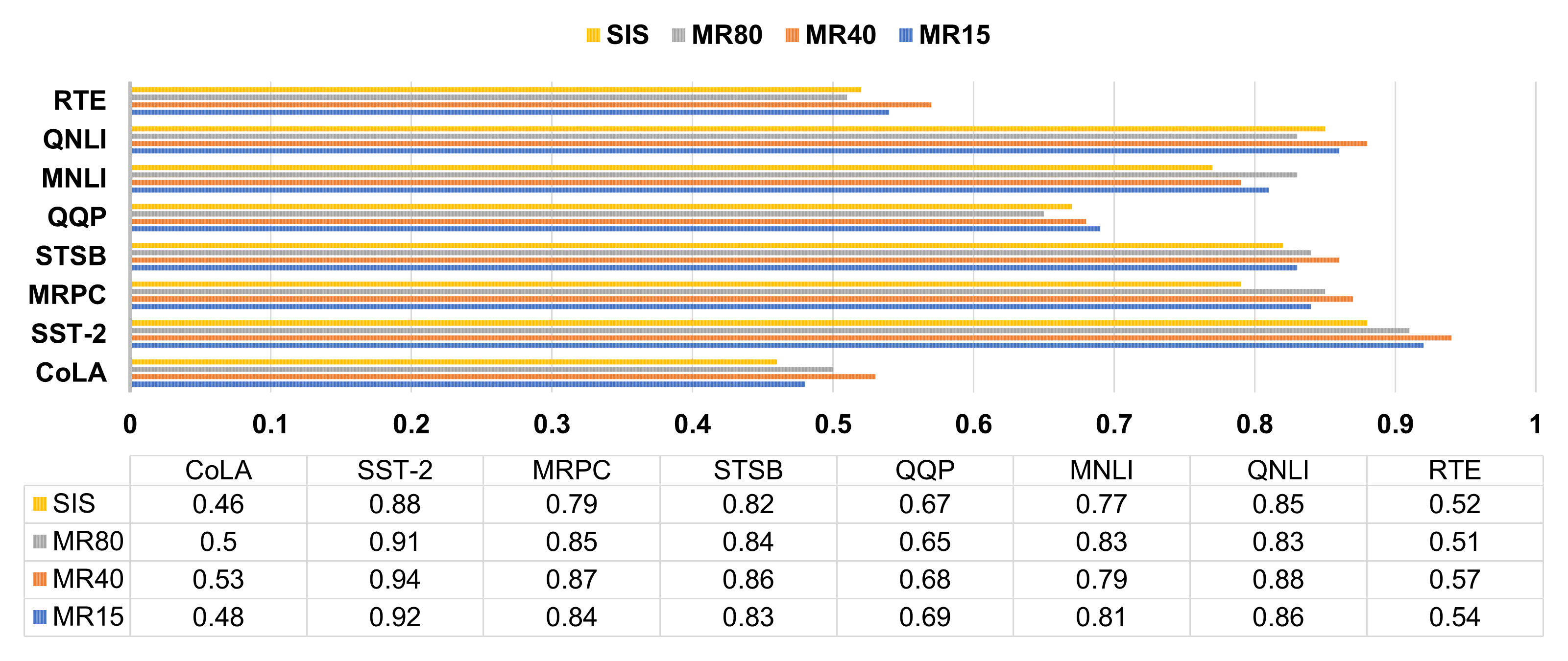}
    \caption{Performance of GPT under different masking rates and SIS. ``Orange" represents the results of the SIS trained model, ``Blue" represents the results of the 15\% masking rate trained model, and similarly, ``Red" and ``Gray" represent the results of 40\% and 80\% masking rate, respectively.}
    \label{fig:GPT}
    \vspace{-0.4cm}
\end{figure*}

We conduct two types of experiments using the GPT model, with different masking rates (15\% masking rate, 40\% masking rate, and 80\% masking rate) and SIS. Fig. \ref{fig:GPT} depicts the metrics of GPT under different masking rates and SIS. In terms of overall performance, the 15\% masking rate performs the best and achieves better results in all tasks. For the GPT model, the best-performing task with a masking rate of 15\% is SST-2, which can achieve 92\%; the worst-performing task is CoLA, with 48\%. On the other hand, the performance of the model trained with 40\% masking improves in most tasks, such as the SST-2 task and the CoLA task. When the masking rate is increased to 80\%, most of the inputs have been masked, and the performance of the model decreases somewhat, but not significantly. Meanwhile, the results show that the performance does not decrease but improves as the masking rate increases. In the case of high masking rates, the GPT can still make effective judgments, and the results do not differ much between different masking rates. After training the model with SIS, the model is still able to perform each task effectively, with results that vary within 10\% from the other models. This result proves the effectiveness of the sufficient input subset algorithm. It also indicates that the GPT model produces an overinterpretation phenomenon of the input dataset. The data contained in the SIS is extremely sparse, and some words are not associated at all. Therefore, there is an overinterpretation phenomenon for the dataset by using GPT.

\subsection{BERT}
In October 2018, Google AI published their BERT \cite{DevlinJacob2018BERT}, a pre-trained language representation model. It emphasizes that BERT uses the new masked language model (MLM) so that deep bidirectional language representations can be generated. The pre-trained BERT model can be fine-tuned with just one additional output layer to create state-of-the-art models for a wide range of tasks, such as question answering and language inference, without substantial task-specific architecture modifications. 

\begin{figure}
    \centering
    \includegraphics[scale=0.6]{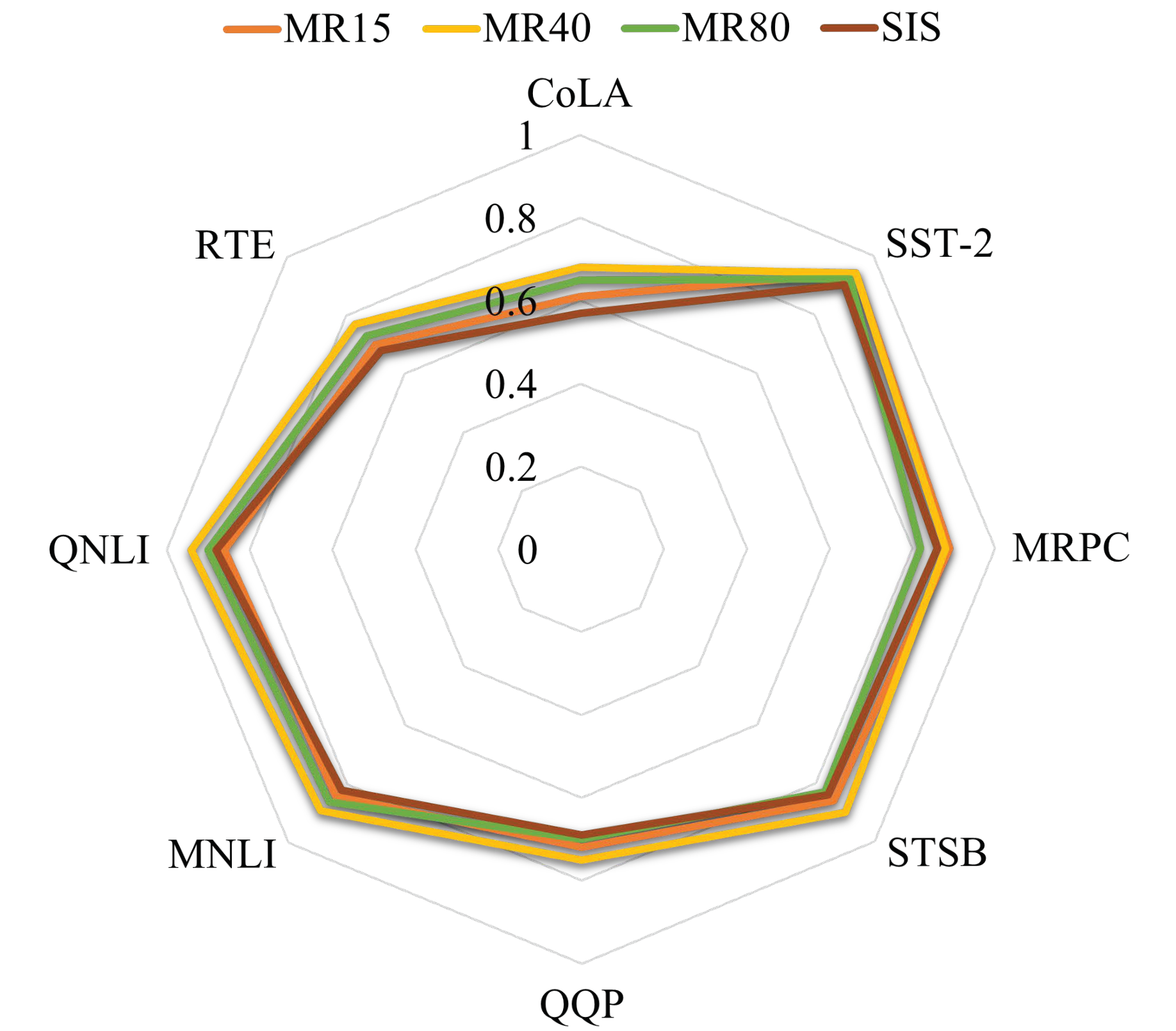}
    \caption{Performance of BERT. ``Orange" represents the performance in a 15\% masking rate strategy. ``Gold" represents the performance at 40\% masking rate strategy. ``Green" denotes the performance of BERT under 80\% masking rate. ``Brown'' shows the performance of the SIS-trained BERT.}
    \label{fig:bert}
\end{figure}

Fig. \ref{fig:bert} shows the results of BERT with different masking rate strategies (15\% masking rate, 40\% masking rate, and 80\% masking rate) and SIS. As shown in Fig. \ref{fig:bert}, the graphs of the 15\% masking rate strategy and the 40\% masking rate strategy are largely merged, and the graph of the 40\% masking rate masks the graph of the 15\% masking rate. This result shows that the performance of the model does not decrease with increasing masking rate, but improves when applying 15\% masking and 40\% masking to the dataset. The performance of the BERT model is improved with reduced input datasets. Moreover, when the masking rate is increased to 80\%, the performance of BERT decreases, but it is still better than the model trained with a 15\% masking rate. The higher masking rate means that the input dataset to the model contains fewer data, and the knowledge learned by the classifier is relatively lower. However, the BERT can still achieve good results. Even when keeping only 20\% of the input dataset, it still outperforms the 15\% masking rate model. To validate the final data or features used by BERT, we extract sufficient input subsets to train the model using the SIS algorithm. In the case of extremely sparse data, BERT can perform each task accurately. This fact shows that the BERT model can learn ``knowledge" that humans cannot understand. Meanwhile, the experimental results show that deep learning is not affected by this aspect and still learns useful information to make final judgments at high masking rates. This result suggests that BERT does not learn real knowledge, but rather statistical signals for the dataset. Therefore, BERT is suffering from overinterpretation.

\subsection{XLNet}
Unlike BERT, XLNet \cite{YangZhilin2019XLNET} is essentially the idea of using an autoregressive language model to encode bi-directional semantic information simultaneously, which can overcome the problems of missing dependencies and inconsistent training/fine-tuning that exist in BERT. Furthermore, XLNet integrates ideas from Transformer-XL, the state-of-the-art autoregressive model, into pre-training. 

\begin{table*}[]
\centering
\caption{XLNet under different masking rates and SIS. ``MR15" represents a 15\% masking rate, and similarly, ``MR40" and ``MR80" represent a 40\% and 80\% masking rate, respectively.}
\label{mask_X}
\begin{tabular}{cccccccccc}
\hline
\multirow{2}{*}{Works} & \multirow{2}{*}{Pre-training} & \multicolumn{8}{c}{Metrics}                             \\ \cline{3-10} 
                       &                               & MNLI & QQP  & QNLI & SST-2 & CoLA & STS-B & MRPC & RTE  \\ \hline
\multirow{4}{*}{XLNet} & MR15                          & 0.90 & 0.92 & 0.94 & 0.97  & 0.69 & 0.92  & 0.90 & 0.85 \\ \cline{2-10} 
                       & MR40                          & 0.93 & 0.93 & 0.92 & 0.95  & 0.73 & 0.94  & 0.95 & 0.89 \\ \cline{2-10} 
                       & MR80                          & 0.91 & 0.92 & 0.91 & 0.94  & 0.70 & 0.93  & 0.92 & 0.87 \\ \cline{2-10} 
                       & SIS                           & 0.86 & 0.88 & 0.90 & 0.94  & 0.66 & 0.88  & 0,85 & 0.81 \\ \hline
\end{tabular}
\end{table*}

Table \ref{mask_X} shows the metrics of XLNet under different training strategies, 15\% masking rate, 40\% masking rate, and 80\% masking rate. XLNet achieves good performance in all eight downstream tasks. By comparing the 15\% masking rate model with the 40\% masking rate model, all downstream tasks improve with increasing masking rate except for QNLI and SST-2. Among them, MRPC shows the most significant improvement at 5\%, with the other tasks showing improvements between 2\% and 3\%. Then, comparing the 80\% masking rate with the 40\% masking rate, the performance of XLNet decreases. However, there are still some improvements in XLNet compared to the 15\% masking rate. The results indicate that training XLNet using the masking rate strategy will further improve the model's capability. It means that XLNet can learn the ``knowledge" to support its final decision after losing 80\% of the input dataset. The remaining 20\% of the dataset is beyond human comprehension, let alone using it to make some series of decisions, such as searching for codes, etc. The performance of XLNet trained with SIS decreases in all tasks, but it is close to the other masking rate models, i.e., results stay within 10\%. This result proves that the SIS extracted by the SIS algorithm is valid. Meanwhile, the SIS contains much less data than the full dataset and lacks many salient features. The XLNet, however, can make accurate decisions from these data. XLNet has its own unique way of learning to understand the dataset, and it is the unknown way of understanding that leads to the overinterpretation of XLNet.

\begin{tcolorbox}[
  colframe = gray!30!white, colback = gray!10!white,
  colbacktitle = gray!30!white,
  coltext = black!50!black,
  coltitle = black!90!white]
{\setlength{\parindent}{0cm}
\textbf{Finding 2:} Overinterpretation is prevalent in PLMs. PLMs trained in different ways behave similarly, where the input varies widely, from a few characters to full queries.}
\end{tcolorbox}

\section{Impact and Mitigation}
\label{sec_miti}
\begin{tcolorbox}[title = {RQ 3:},
  colframe = gray!30!white, colback = gray!10!white,
  colbacktitle = gray!30!white,
  coltext = black!50!black,
  coltitle = black!90!white]
  What is the impact of overinterpretation? What are the challenges in mitigating overinterpretation in general and how to mitigate overinterpretation?  
\end{tcolorbox}

The above experiments show that overinterpretation appears not only in SE tasks but also in PLMs. The essence of both types of experiments is that the input size is reduced and the final result is consistent with the full input. This fact suggests that the presence of overinterpretation allows PLMs to improve the efficiency of training by reducing the input. However, this model, which achieves ultra-high results on a very small set of data, is wrong \cite{CarterBrandon2020Overinterpretation}. The drawback is that PLMs learn contents that are incomprehensible from a human perspective, or more accurately they learn unique statistical signals in the dataset. Moreover, if model decisions are made based on statistical signals alone they can have serious consequences in terms of misclassification. PLMs make incorrect decisions when different datasets produce the same statistical signal. For example, the results of code search do not match and delay software development; the generated code summaries are inaccurate and increase the cost of software post-maintenance; duplicate bug reports are retained or new bug reports are misclassified as duplicates, resulting in bugs that cannot be fixed in time.
Thus, we design a survey to collect developers' perceptions about overinterpretation. The survey contains the following questions: 1) \textit{Do you frequently use SE techniques (such as code search, code summarization, duplicate bug report detection, etc.) in developing and maintaining software?} 2) \textit{Have these techniques ever resulted in errors with serious consequences?} 3) \textit{Can you understand the real meaning of the query after masking?} 4) \textit{In the case of code search, which code fragment is actually queried by the text ``L the ON" after masking?}

We invite five experienced developers who have more than five-years experience on software development and maintenance experience to participate in our survey. For the first question, all five developers use code search frequently to help with development. For the second problem, errors can occur that cause the code to run incorrectly and slow down the software development. For the third problem, all five developers can not understand the meaning of the masked query. For the fourth question, all developers do not find the correct code snippet in the given example. Therefore, we believe that overinterpretation is a serious problem to produce bad influence on the software development and maintenance process.

The following issues affect the mitigation of overinterpretation. First, overinterpretation is not well understood and studied at present. It can be easily misunderstood. Second, overinterpretation is not easily detected. The overinterpretation may arise from the true statistical signal in the underlying dataset distribution. Thus, overinterpretation can be harder to diagnose as it admits decisions that are made by statistically valid criteria, and models that use such criteria can excel at benchmarks. Finally, PLMs are black-box models. Then, we find that the whole word mask strategy and ensembling can mitigate overinterpretation. Both of them can enrich the input dataset. 

\textbf{Whole Word Masking Strategy}: If a subword is masked, the other parts of the same word are also masked, i.e., the whole word masking strategy \cite{CuiYiming2021PWWW}. We describe and summarize the experimental results in this paper.  We still use three mask rates of 15\%, 40\%, and 80\% to train the GPT \cite{Radford2018Corpus}, BERT \cite{DevlinJacob2018BERT}, and XLNet \cite{YangZhilin2019XLNET}. Whole word masking strategy can effectively improve the integrity of the words after masking. Meanwhile, the readability of the text after masking is increased and the number of meaningless words decreases (Table \ref{WWM}). 

\begin{table}[]
\caption{The performance of whole word masking under QNLI.}
\centering
\begin{tabular}{cccc}
\hline
Whole Word Masking & GPT  & BERT & XLNet \\ \hline
15\%               & 0.81 & 0.87 & 0.91  \\ \hline
40\%               & 0.85 & 0.92 & 0.90  \\ \hline
80\%               & 0.83 & 0.89 & 0.87  \\ \hline
\end{tabular}
\label{WWM}
\end{table}

\begin{figure}
    \centering
    \includegraphics[scale=0.4]{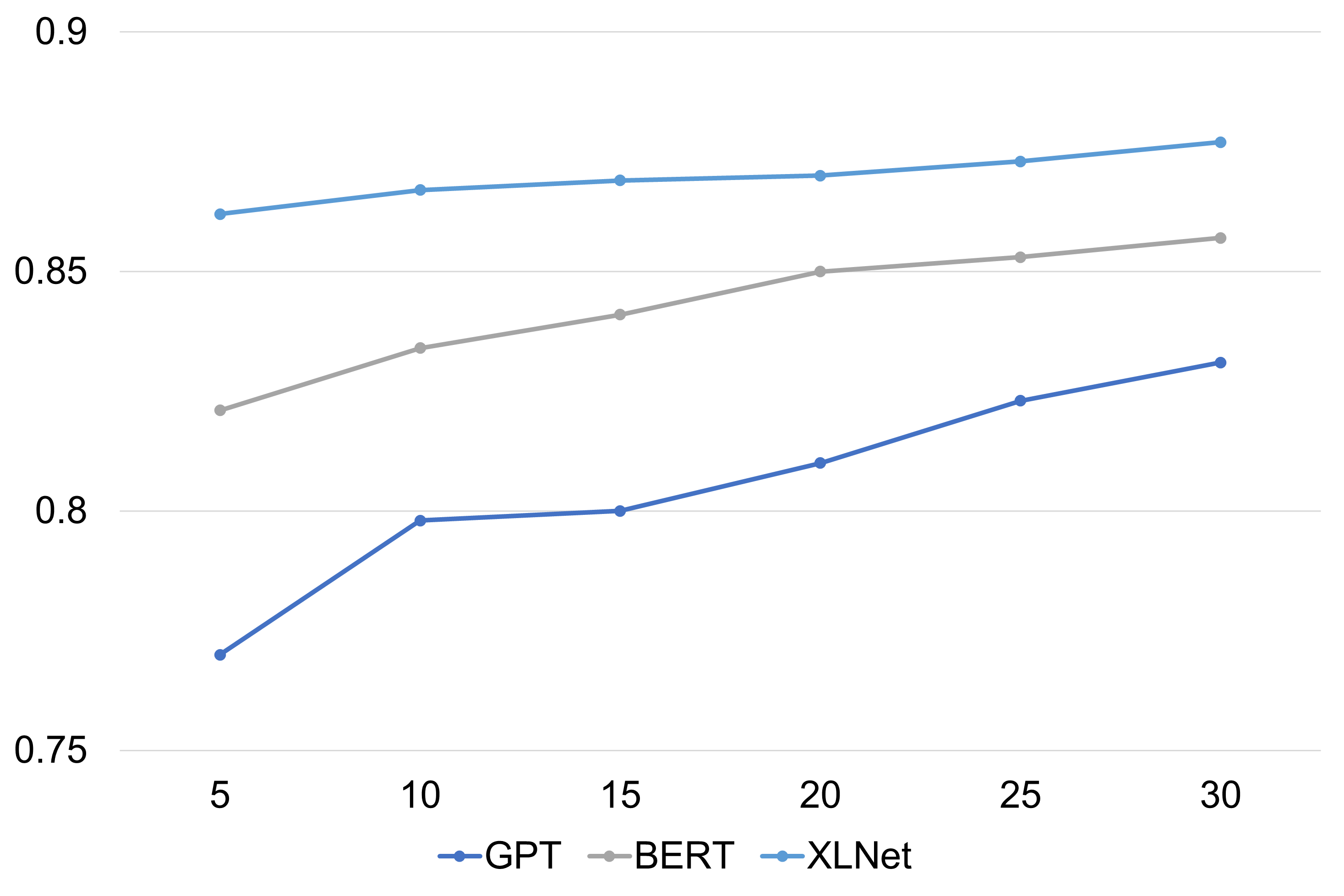}
    \caption{SIS size on MNLI as the number of characters varies. The horizontal axis represents the number of characters contained in the SIS, and the vertical axis represents the MNLI.}
    \label{fig:ensemble}
\end{figure}

\textbf{Ensembling}: It is known to improve classification performance \cite{GohKing-Shy2001SVM, JuCheng2018Trpo}. But it can also be used to increase the SIS size, hence mitigating overinterpretation. We observe that SIS subsets are generally not transferable from one model to another i.e., an SIS for one model is rarely an SIS for another. Thus, different models rely on different independent signals to arrive at the same prediction. We find that ensembling uniformly increases test accuracy as expected but also increases the SIS size (Fig. \ref{fig:ensemble} ).

\begin{tcolorbox}[
  colframe = gray!30!white, colback = gray!10!white,
  colbacktitle = gray!30!white,
  coltext = black!50!black,
  coltitle = black!90!white]
{\setlength{\parindent}{0cm}
\textbf{Finding 3:} Overinterpretation makes the model learn only the statistical signal and ignore the crucial features. Whole word masking and ensembling are found to mitigate overinterpretation.}
\end{tcolorbox}

\section{THREATS TO VALIDITY}
\label{sec_threats}

\textbf{Internal threats to validity.} The inputs used for each scenario and model are different, and we have chosen to use the dataset that they originally used rather than the new uniform dataset. In the future, we will experiment on several different datasets. In this way, we will verify that overinterpretation does not depend on unique datasets.

{\setlength{\parindent}{0cm}
\textbf{External threats to validity.} Our study has a limited scope and only three types of AI4SE tasks (code search \cite{GuXiaodong2018DeepCS,FengZhangyin2020CodeBERT}, code summarization \cite{AhmadWasiUddin2020Transformer,IyerSrinivasan2016Summarizing}, and duplicate bug report detection \cite{XiaoGuanping2020HINDBR,BudhirajaAmar2018DWEN}) and three pre-trained language models (GPT \cite{Radford2018Corpus}, BERT \cite{DevlinJacob2018BERT}, and XLNet \cite{YangZhilin2019XLNET}) have been selected. These tasks are only a part of the SE tasks. There are many different types of tasks that have not been studied and demonstrated for overinterpretation. In the future, we will select more various AI4SE tasks and models to validate our scheme.}

\section{RELATED WORK}
\label{sec_rw}
\textbf{AI4SE} In recent years, many empirical studies have focused on AI and SE tasks. But these researches are limited to only one aspect. Wu et al. \cite{WuYanzhao2022ACMS} conduct an empirical comparison and analysis of four representative deep learning frameworks with three unique contributions. Christian et al. \cite{GarbinChristian2020Dvbn} conduct an empirical study to investigate the effect of dropout and batch normalization on training deep learning models. Hu et al. \cite{HuQiang2022AESo} first conduct a systemically empirical study to reveal the impact of the retraining process and data distribution on model enhancement. Abdallah et al. \cite{SemasabaAbubakarOmariAbdallah2022Aeeo} propose an evaluation of vulnerability detection performance on source code representations and evaluates how DL strategies can improve them. Du et al. \cite{DuXiaoting2022AESo} present the first comprehensive empirical study on fault triggering conditions in three widely-used deep learning frameworks. Pan et al. \cite{PanCong2021Aeso} perform empirical studies using DL models in cross-version and cross-project software defect prediction to investigate if using a neural language model could improve prediction performance. However, all these studies failed to examine the flaws of deep learning models themselves.

Meanwhile, many researchers have started to focus on explainable AI for SE tasks. Rabin et al. \cite{RabinMdRafiqulIslam2021UNCI} propose a model-agnostic approach to identify critical input features for models in code intelligence tools, by drawing on software debugging research, and then exploring and analyzing the models. Cito et al. \cite{CitoJurgen2022CEfM} explore counterfactual explanations for models of source code to help developers understand and use the model. Li et al. \cite{LiYi2021VDwF} propose IVDetect, an interpretable vulnerability detector that uses AI to detect vulnerabilities while providing an interpretation of the vulnerability detector. These works are concerned with how to interpret the decisions made by the model. However, we explore a potential flaw of the model, i.e. overinterpretation.

{\setlength{\parindent}{0cm}\textbf{PLMs} Previous studies study cross-modal pre-trained language models \cite{ZengZhixiong2022ACES} and the robustness of pre-trained language models to spurious correlations \cite{TuLifu2020AESo}. Moreover, many studies focus on the design of pre-trained models and the enhancement of models \cite{VaswaniAshish2017Aiay}. The first generation pretrained models aim to learn good word embeddings, they are usually very shallow for computational efficiencies, such as Skip-Gram \cite{MikolovTomas2013DRoW} and GloVe \cite{PenningtonJeffrey2014GGvf}. The second generation pre-trained models focus on learning contextual word embeddings, such as CoVe \cite{McCannBryan2017LTCWV}, ELMo \cite{PetersMatthewE2018Dcwr}. All the above works do not focus on some defects hidden by the models, such as overinterpretation. This study explains overinterpretation by investigating different pre-trained language models.}

\section{CONCLUSION}
\label{sec_conclu}

Deep learning-based natural language processing techniques are becoming increasingly popular for researchers to solve various tasks in SE. This paper constructs the first comprehensive study to reveal the overinterpretation in PLMs of AI4SE tasks. By investigating the most representative AI4SE tasks as well as PLMs, we identify the existence of overinterpretation in these models. The wide presence of these problems motivates future research to further tackle the overinterpretation of deep learning. 

In the future, we will design a new evaluation scheme to identify overinterpretation and thus help researchers refine their models.

\section*{Acknowledgments}

The authors would like to appreciate all anonymous reviewers for their insightful comments and constructive suggestions to polish this paper in high quality.

\bibliographystyle{IEEEtran}
\bibliography{reference} 

\begin{thebibliography}{10}
\providecommand{\url}[1]{#1}
\csname url@samestyle\endcsname
\providecommand{\newblock}{\relax}
\providecommand{\bibinfo}[2]{#2}
\providecommand{\BIBentrySTDinterwordspacing}{\spaceskip=0pt\relax}
\providecommand{\BIBentryALTinterwordstretchfactor}{4}
\providecommand{\BIBentryALTinterwordspacing}{\spaceskip=\fontdimen2\font plus
\BIBentryALTinterwordstretchfactor\fontdimen3\font minus
  \fontdimen4\font\relax}
\providecommand{\BIBforeignlanguage}[2]{{%
\expandafter\ifx\csname l@#1\endcsname\relax
\typeout{** WARNING: IEEEtran.bst: No hyphenation pattern has been}%
\typeout{** loaded for the language `#1'. Using the pattern for}%
\typeout{** the default language instead.}%
\else
\language=\csname l@#1\endcsname
\fi
#2}}
\providecommand{\BIBdecl}{\relax}
\BIBdecl

\bibitem{GeziciBahar2022Systematic}
B.~Gezici and A.~K. Tarhan, ``Systematic literature review on software quality
  for ai-based software,'' \emph{Empirical software engineering : an
  international journal}, vol.~27, pp. 1--65, 2022.

\bibitem{HuangAiMing2014Study}
A.~M. Huang, G.~S. Deng, J.~Hu, and Y.~J. Huang, ``Study on cmm-based software
  quality assurance process improvement - a case of the educational software
  quality assurance model,'' vol. 1049, 2014, pp. 2032--2036.

\bibitem{ChenShih-Yeh2022Exploring}
S.-Y. Chen, Y.-S. Su, Y.-Y. Ku, C.-F. Lai, and K.-L. Hsiao, ``Exploring the
  factors of students' intention to participate in ai software development,''
  \emph{Library hi tech}, pp. 1--17, 2022.

\bibitem{Palomo-DuarteManuel2021Assessment}
``Assessment in software development for competitive environments: An ai
  strategy development case study,'' \emph{Electronics (Basel)}, vol.~10, pp.
  1566--1584, 2021.

\bibitem{SheorajYugeshwaree2022UsingAI}
Y.~Sheoraj and R.~K. Sungkur, ``Using ai to develop a framework to prevent
  employees from missing project deadlines in software projects - case study of
  a global human capital management (hcm) software company,'' \emph{Advances in
  engineering software}, vol. 170, pp. 103\,143--103\,156, 2022.

\bibitem{ZhengZhi-Ming2008Earned}
Z.-M. Zheng and A.-H. Ren, ``Earned value method and application in software
  project management,'' \emph{Computer Engineering and Design}, vol.~29, pp.
  4302--4304, 2008.

\bibitem{EspinozaGabrielZ2021EDLm}
G.~Z. Espinoza, R.~M. Angelo, P.~R. Oliveira, and K.~M. Honorio, ``Evaluating
  deep learning models for predicting alk-5 inhibition,'' \emph{PloS one},
  vol.~16, pp. 1--16, 2021.

\bibitem{QuYu2021Evaluating}
Y.~Qu and H.~Yin, ``Evaluating network embedding techniques' performances in
  software bug prediction,'' \emph{Empirical software engineering : an
  international journal}, vol.~26, pp. 60--104, 2021.

\bibitem{DevlinJacob2018BERT}
J.~Devlin, M.-W. Chang, K.~Lee, and K.~Toutanova, ``Bert: Pre-training of deep
  bidirectional transformers for language understanding,'' pp. 1--16, 2018.

\bibitem{CarterBrandon2020Overinterpretation}
B.~Carter, S.~Jain, J.~W. Mueller, and D.~Gifford, ``Overinterpretation reveals
  image classification model pathologies,'' in \emph{Advances in Neural
  Information Processing Systems}, vol.~34.\hskip 1em plus 0.5em minus
  0.4em\relax Curran Associates, 2021, pp. 15\,395--15\,407.

\bibitem{Chakraborty2022TSE}
S.~Chakraborty, R.~Krishna, Y.~Ding, and B.~Ray, ``Deep learning based
  vulnerability detection: Are we there yet?'' \emph{IEEE Transactions on
  Software Engineering}, vol.~48, pp. 3280--3296, 2022.

\bibitem{RibeiroMarcoTulio2016"SIT}
M.~T. Ribeiro, S.~Singh, and C.~Guestrin, ``"why should i trust you?":
  Explaining the predictions of any classifier,'' in \emph{ACM SIGKDD
  International Conference on Knowledge Discovery and Data Mining}, 2016, pp.
  1135--1144.

\bibitem{GuXiaodong2018DeepCS}
``Deep code search,'' in \emph{IEEE/ACM International Conference on Software
  Engineering}, 2018, pp. 933--944.

\bibitem{FengZhangyin2020CodeBERT}
Z.~Feng, D.~Guo, D.~Tang, N.~Duan, X.~Feng, M.~Gong, L.~Shou, B.~Qin, T.~Liu,
  D.~Jiang, and M.~Zhou, ``Codebert: A pre-trained model for programming and
  natural languages,'' in \emph{Findings of the Association for Computational
  Linguistics}, 2020, pp. 1536--1547.

\bibitem{AhmadWasiUddin2020Transformer}
W.~U. Ahmad, S.~Chakraborty, B.~Ray, and K.-W. Chang, ``A transformer-based
  approach for source code summarization,'' pp. 1--10, 2020.

\bibitem{IyerSrinivasan2016Summarizing}
S.~Iyer, I.~Konstas, A.~Cheung, and L.~Zettlemoyer, ``Summarizing source code
  using a neural attention model,'' in \emph{Annual Meeting of the Association
  for Computational Linguistics}, vol.~4, 2016, pp. 2073--2083.

\bibitem{XiaoGuanping2020HINDBR}
G.~Xiao, X.~Du, Y.~Sui, and T.~Yue, ``Hindbr: Heterogeneous information network
  based duplicate bug report prediction,'' in \emph{IEEE International
  Symposium on Software Reliability Engineering}, vol. 2020, 2020, pp.
  195--206.

\bibitem{BudhirajaAmar2018DWEN}
``Poster: Dwen: Deep word embedding network for duplicate bug report detection
  in software repositories,'' vol. 137351, 2018, pp. 193--194.

\bibitem{Radford2018Corpus}
A.~Radford, K.~Narasimhan, T.~Salimans, and I.~Sutskever, ``Improving language
  understanding by generative pre-training,'' pp. 1--12, 2018.

\bibitem{YangZhilin2019XLNET}
Z.~Yang, Z.~Dai, Y.~Yang, J.~Carbonell, R.~Salakhutdinov, and Q.~V. Le,
  ``Xlnet: Generalized autoregressive pretraining for language understanding,''
  in \emph{Neural Information Processing Systems}, 2019, pp. 1--18.

\bibitem{SridharaGiriprasad2010Tags}
G.~Sridhara, E.~Hill, D.~Muppaneni, L.~Pollock, and K.~Vijay-Shanker, ``Towards
  automatically generating summary comments for java methods,'' in
  \emph{Proceedings of the IEEE/ACM International Conference on Automated
  Software Engineering}, 2010, pp. 43--52.

\bibitem{PollockLori2013NLSA}
L.~Pollock, K.~Vijay-Shanker, E.~Hill, G.~Sridhara, and D.~Shepherd, ``Natural
  language-based software analyses and tools for software maintenance,'' in
  \emph{Lecture Notes in Computer Science}, 2013, vol. 7171, pp. 94--125.

\bibitem{FowkesJaroslav2017AfSC}
``Autofolding for source code summarization,'' \emph{IEEE transactions on
  software engineering}, vol.~43, pp. 1095--1109, 2017.

\bibitem{HintonGeoffreyE2006Fast}
G.~E. Hinton, S.~Osindero, and Y.-W. Teh, ``A fast learning algorithm for deep
  belief nets,'' \emph{Neural computation}, vol.~18, pp. 1527--1554, 2006.

\bibitem{KanekoMasahiro2020Encoder}
M.~Kaneko, M.~Mita, S.~Kiyono, J.~Suzuki, and K.~Inui, ``Encoder-decoder models
  can benefit from pre-trained masked language models in grammatical error
  correction,'' in \emph{Annual Meeting of the Association for Computational
  Linguistics}, 2020, pp. 4248--4254.

\bibitem{LabehatKryeziu2022ASoU}
L.~Kryeziu and V.~Shehu, ``A survey of using unsupervised learning techniques
  in building masked language models for low resource languages,'' in \emph{The
  Institute of Electrical and Electronics Engineers}, 2022.

\bibitem{LiuYinhan2019RARO}
Y.~Liu, M.~Ott, N.~Goyal, J.~Du, M.~Joshi, D.~Chen, O.~Levy, M.~Lewis,
  L.~Zettlemoyer, and V.~Stoyanov, ``Roberta: A robustly optimized bert
  pretraining approach,'' pp. 1--13, 2019.

\bibitem{LevineYoav2020PPmo}
Y.~Levine, B.~Lenz, O.~Lieber, O.~Abend, K.~Leyton-Brown, M.~Tennenholtz, and
  Y.~Shoham, ``Pmi-masking: Principled masking of correlated spans,'' pp.
  1--13, 2020.

\bibitem{Perez2022SD}
J.~Perez, J.~L. Flores, C.~Blum, J.~Cerquides, and A.~Abuin, ``Optimization
  techniques and formal verification for the software design of boolean algebra
  based safety-critical systems,'' \emph{IEEE Transactions on Industrial
  Informatics}, vol.~18, pp. 620--630, 2022.

\bibitem{Maletic1994M}
J.~Maletic and R.~Reynolds, ``A tool to support knowledge based software
  maintenance: the software service bay,'' in \emph{International Conference on
  Tools with Artificial Intelligence.}, 1994, pp. 11--17.

\bibitem{NiuHaoran2016Ltrc}
H.~Niu, I.~Keivanloo, and Y.~Zou, ``Learning to rank code examples for code
  search engines,'' \emph{Empirical software engineering : an international
  journal}, vol.~22, pp. 259--291, 2016.

\bibitem{RaghothamanMukund2016SWIM}
M.~Raghothaman, Y.~Wei, and Y.~Hamadi, ``Swim: Synthesizing what i mean - code
  search and idiomatic snippet synthesis,'' in \emph{IEEE/ACM International
  Conference on Software Engineering}, vol.~14, 2016, pp. 357--367.

\bibitem{LiXuan2016Relationship-Aware}
X.~Li, Z.~Wang, Q.~Wang, S.~Yan, T.~Xie, and H.~Mei, ``Relationship-aware code
  search for javascript frameworks,'' in \emph{ACM SIGSOFT International
  Symposium on Foundations of Software Engineering}, 2016, pp. 690--701.

\bibitem{LvFei2015CodeHow}
F.~Lv, H.~Zhang, J.-G. Lou, S.~Wang, D.~Zhang, and J.~Zhao, ``Codehow:
  Effective code search based on api understanding and extended boolean model
  (e),'' in \emph{IEEE/ACM International Conference on Automated Software
  Engineering}, 2015, pp. 260--270.

\bibitem{McMillanCollin2011Pfrf}
C.~McMillan, M.~Grechanik, D.~Poshyvanyk, Q.~Xie, and C.~Fu, ``Portfolio:
  finding relevant functions and their usage,'' in \emph{International
  Conference on Software Engineering}, 2011, pp. 111--120.

\bibitem{PonzanelliLuca2014Mstt}
L.~Ponzanelli, G.~Bavota, M.~Di~Penta, R.~Oliveto, and M.~Lanza, ``Mining
  stackoverflow to turn the ide into a self-confident programming prompter,''
  in \emph{Working Conference on Mining Software Repositories}, 2014, pp.
  102--111.

\bibitem{RobertsonStephen2004SBet}
S.~Robertson, H.~Zaragoza, and M.~Taylor, ``Simple bm25 extension to multiple
  weighted fields,'' in \emph{International Conference on Information and
  Knowledge Management}, 2004, pp. 42--49.

\bibitem{SunChengnian2011Tmar}
C.~Sun, D.~Lo, S.-C. Khoo, and J.~Jiang, ``Towards more accurate retrieval of
  duplicate bug reports,'' in \emph{IEEE/ACM International Conference on
  Automated Software Engineering}, 2011, pp. 253--262.

\bibitem{DeshmukhJayati2017TADB}
J.~Deshmukh, K.~M. Annervaz, S.~Podder, S.~Sengupta, and N.~Dubash, ``Towards
  accurate duplicate bug retrieval using deep learning techniques,'' in
  \emph{IEEE International Conference on Software Maintenance and Evolution},
  2017, pp. 115--124.

\bibitem{AbbesM2011AESo}
M.~Abbes, F.~Khomh, Y.-G. Guéhéneuc, and G.~Antoniol, ``An empirical study
  of the impact of two antipatterns, blob and spaghetti code, on program
  comprehension,'' in \emph{European Conference on Software Maintenance and
  Reengineering}, 2011, pp. 181--190.

\bibitem{PolitowskiCristiano2020Alse}
``A large scale empirical study of the impact of spaghetti code and blob
  anti-patterns on program comprehension,'' \emph{Information and software
  technology}, vol. 122, pp. 106\,278--106\,292, 2020.

\bibitem{OlbrichS2009Teai}
S.~Olbrich, D.~Cruzes, V.~Basili, and N.~Zazworka, ``The evolution and impact
  of code smells: A case study of two open source systems,'' in
  \emph{International Symposium on Empirical Software Engineering and
  Measurement}, 2009, pp. 390--400.

\bibitem{CornelissenB2009ASSo}
``A systematic survey of program comprehension through dynamic analysis,''
  \emph{IEEE transactions on software engineering}, vol.~35, pp. 684--702,
  2009.

\bibitem{NoughiNesrine2017AESo}
N.~Noughi, S.~Hanenberg, and A.~Cleve, ``An empirical study on the usage of sql
  execution traces for program comprehension,'' in \emph{IEEE International
  Conference on Software Quality, Reliability and Security Companion}, 2017,
  pp. 47--54.

\bibitem{YanShuhan2020AtCS}
``Are the code snippets what we are searching for? a benchmark and an empirical
  study on code search with natural-language queries,'' in \emph{IEEE
  International Conference on Software Analysis, Evolution and Reengineering},
  2020, pp. 344--354.

\bibitem{McBurneyPaulW2014Aeso}
P.~W. McBurney and C.~McMillan, ``An empirical study of the textual similarity
  between source code and source code summaries,'' \emph{Empirical software
  engineering : an international journal}, vol.~21, pp. 17--42, 2014.

\bibitem{HaoRui2022Adrh}
R.~Hao, Y.~Li, Y.~Feng, and Z.~Chen, ``Are duplicates really harmful? an
  empirical study on bug report summarization techniques,'' \emph{Journal of
  software : evolution and process}, 2022.

\bibitem{CarterBrandon2018Black-box}
B.~Carter, J.~Mueller, S.~Jain, and D.~Gifford, ``What made you do this?
  understanding black-box decisions with sufficient input subsets,'' vol.~89,
  pp. 1--35, 2018.

\bibitem{SafdariNasir2018LtRR}
N.~Safdari, ``Learning to rank relevant files for bug reports using domain
  knowledge, replication and extension of a learning-to-rank approach,'' pp.
  1--53, 2018.

\bibitem{PapineniKishore2002BLEU}
K.~Papineni, S.~Roukos, T.~Ward, and W.-J. Zhu, ``Bleu: A method for automatic
  evaluation of machine translation,'' in \emph{Annual Meeting on Association
  for Computational Linguistics}, 2002, pp. 311--318.

\bibitem{BanerjeeSatanjeev2005Meteor}
S.~Banerjee and A.~Lavie, ``Meteor: An automatic metric for mt evaluation with
  improved correlation with human judgments,'' in \emph{Intrinsic and Extrinsic
  Evaluation Measures for Machine Translation and/or Summarization}, 2005, pp.
  65--72.

\bibitem{lin2004rouge}
C.-Y. Lin, ``Rouge: a package for automatic evaluation of summaries,'' in
  \emph{Workshop on Text Summarization Branches Out, Post-Conference Workshop
  of ACL}, 2004, pp. 74--81.

\bibitem{NangiaNikita2017TR2S}
N.~Nangia, A.~Williams, A.~Lazaridou, and S.~R. Bowman, ``The repeval 2017
  shared task: Multi-genre natural language inference with sentence
  representations,'' pp. 1--10, 2017.

\bibitem{RajpurkarPranav2016SQuAD}
P.~Rajpurkar, J.~Zhang, K.~Lopyrev, and P.~Liang, \emph{SQuAD: 100,000+
  Questions for Machine Comprehension of Text}, 2016.

\bibitem{Warstadt2018Corpus}
\BIBentryALTinterwordspacing
A.~Warstadt, A.~Singh, and S.~R. Bowman, \emph{Corpus of linguistic
  acceptability}, 2018. [Online]. Available:
  \url{http://nyu-mll.github.io/cola}
\BIBentrySTDinterwordspacing

\bibitem{SocherRichard2013Recursive}
R.~Socher, A.~Perelygin, J.~Wu, J.~Chuang, C.~Manning, A.~Ng, and C.~Potts,
  ``Recursive deep models for semantic compositionality over a sentiment
  treebank,'' \emph{EMNLP}, vol. 1631, pp. 1631--1642, 2013.

\bibitem{Dolan2022Automatically}
W.~Dolan and C.~Brockett, ``Automatically constructing a corpus of sentential
  paraphrases,'' in \emph{International Workshop on Paraphrasing}, 2022, pp.
  1--8.

\bibitem{Chen2018Quora}
\BIBentryALTinterwordspacing
Z.~Chen, H.~Zhang, X.~Zhang, and L.~Zhao, \emph{Quora question pairs}, 2018.
  [Online]. Available:
  \url{https://data.quora.com/First-QuoraDataset-Release-Questio -Pairs}
\BIBentrySTDinterwordspacing

\bibitem{CerD2017ST1s}
D.~Cer, M.~Diab, E.~Agirre, I.~Lopez-Gazpio, and L.~Specia, ``Semeval-2017 task
  1: semantic textual similarity - multilingual and cross-lingual focused
  evaluation.''\hskip 1em plus 0.5em minus 0.4em\relax Association for
  Computational Linguistics, 2017.

\bibitem{CuiYiming2021PWWW}
Y.~Cui, W.~Che, T.~Liu, B.~Qin, and Z.~Yang, ``Pre-training with whole word
  masking for chinese bert,'' \emph{IEEE/ACM transactions on audio, speech, and
  language processing}, vol.~29, pp. 3504--3514, 2021.

\bibitem{GohKing-Shy2001SVM}
K.-S. Goh, E.~Chang, and K.-T. Cheng, ``Svm binary classifier ensembles for
  image classification,'' in \emph{International Conference on Information and
  Knowledge Management, Proceedings}, 2001, pp. 395--402.

\bibitem{JuCheng2018Trpo}
C.~Ju, A.~Bibaut, and M.~van~der Laan, ``The relative performance of ensemble
  methods with deep convolutional neural networks for image classification,''
  \emph{Journal of applied statistics}, vol.~45, pp. 2800--2818, 2018.

\bibitem{WuYanzhao2022ACMS}
Y.~Wu, L.~Liu, C.~Pu, W.~Cao, S.~Sahin, W.~Wei, and Q.~Zhang, ``A comparative
  measurement study of deep learning as a service framework,'' \emph{IEEE
  transactions on services computing}, vol.~15, pp. 551--566, 2022.

\bibitem{GarbinChristian2020Dvbn}
C.~Garbin, X.~Zhu, and O.~Marques, ``Dropout vs. batch normalization: an
  empirical study of their impact to deep learning,'' \emph{Multimedia tools
  and applications}, vol.~79, pp. 12\,777--12\,815, 2020.

\bibitem{HuQiang2022AESo}
Q.~Hu, Y.~Guo, M.~Cordy, X.~Xie, L.~Ma, M.~Papadakis, and Y.~Le~Traon, ``An
  empirical study on data distribution-aware test selection for deep learning
  enhancement,'' \emph{ACM transactions on software engineering and
  methodology}, 2022.

\bibitem{SemasabaAbubakarOmariAbdallah2022Aeeo}
A.~O.~A. Semasaba, W.~Zheng, X.~Wu, S.~A. Agyemang, T.~Liu, and Y.~Ge, ``An
  empirical evaluation of deep learning-based source code vulnerability
  detection: Representation versus models,'' \emph{Journal of software :
  evolution and process}, pp. 1--23, 2022.

\bibitem{DuXiaoting2022AESo}
X.~Du, Y.~Sui, Z.~Liu, and J.~Ai, ``An empirical study of fault triggers in
  deep learning frameworks,'' \emph{IEEE transactions on dependable and secure
  computing}, pp. 1--18, 2022.

\bibitem{PanCong2021Aeso}
C.~Pan, M.~Lu, and B.~Xu, ``An empirical study on software defect prediction
  using codebert model,'' \emph{Applied sciences}, vol.~11, pp. 4793--4813,
  2021.

\bibitem{RabinMdRafiqulIslam2021UNCI}
M.~R.~I. Rabin, V.~J. Hellendoorn, and M.~A. Alipour, ``Understanding neural
  code intelligence through program simplification,'' 2021, pp. 441--452.

\bibitem{CitoJurgen2022CEfM}
J.~Cito, I.~Dillig, V.~Murali, and S.~Chandra, ``Counterfactual explanations
  for models of code,'' in \emph{IEEE/ACM International Conference on Software
  Engineering: Software Engineering in Practice}, 2022, pp. 125--134.

\bibitem{LiYi2021VDwF}
Y.~Li, S.~Wang, and T.~N. Nguyen, ``Vulnerability detection with fine-grained
  interpretations,'' 2021, pp. 292--303.

\bibitem{ZengZhixiong2022ACES}
Z.~Zeng and W.~Mao, ``A comprehensive empirical study of vision-language
  pre-trained model for supervised cross-modal retrieval,'' pp. 1--11, 2022.

\bibitem{TuLifu2020AESo}
L.~Tu, G.~Lalwani, S.~Gella, and H.~He, ``An empirical study on robustness to
  spurious correlations using pre-trained language models,'' \emph{Transactions
  of the Association for Computational Linguistics}, vol.~8, pp. 621--633,
  2020.

\bibitem{VaswaniAshish2017Aiay}
A.~Vaswani, N.~Shazeer, N.~Parmar, J.~Uszkoreit, L.~Jones, A.~N. Gomez,
  Å.~Kaiser, and I.~Polosukhin, ``Attention is all you need,'' in
  \emph{Advances in Neural Information Processing Systems}, vol. 2017, 2017,
  pp. 5999--6009.

\bibitem{MikolovTomas2013DRoW}
T.~Mikolov, I.~Sutskever, K.~Chen, G.~Corrado, and J.~Dean, ``Distributed
  representations of words and phrases and their compositionality,'' pp. 1--9,
  2013.

\bibitem{PenningtonJeffrey2014GGvf}
J.~Pennington, R.~Socher, and C.~D. Manning, ``Glove: Global vectors for word
  representation,'' in \emph{Conference on Empirical Methods in Natural
  Language Processing}, 2014, pp. 1532--1543.

\bibitem{McCannBryan2017LTCWV}
B.~McCann, J.~Bradbury, C.~Xiong, and R.~Socher, ``Learned in translation:
  Contextualized word vectors,'' in \emph{Neural Information Processing
  Systems}, 2017, pp. 6297--6308.

\bibitem{PetersMatthewE2018Dcwr}
M.~E. Peters, M.~Neumann, M.~Iyyer, M.~Gardner, C.~Clark, K.~Lee, and
  L.~Zettlemoyer, ``Deep contextualized word representations,'' in
  \emph{Conference of the North American Chapter of the Association for
  Computational Linguistics: Human Language Technologies}, vol.~1, 2018, pp.
  2227--2237.

\end{thebibliography}

\end{document}